\documentclass[fleqn,usenatbib,useAMS]{mnras}

\usepackage{graphicx}
\usepackage{amsmath}
\usepackage{amssymb}
\usepackage{bm}
\usepackage{booktabs}
\usepackage{xcolor}
\usepackage[T1]{fontenc}
\usepackage{newtxtext,newtxmath}
\hypersetup{hypertexnames=false,colorlinks=true,linkcolor=blue,citecolor=blue,urlcolor=blue}

\title[A slim disc-corona model for TDEs]{A magnetically heated slim disc-corona model for tidal disruption events}

\author[Chen et al.]{
Wei Chen$^{1,2,3}$,
Chris Done$^{2}$\thanks{Corresponding author. E-mail: chris.done@durham.ac.uk},
Erlin Qiao$^{1,3}$\thanks{Corresponding author. E-mail: qiaoel@nao.cas.cn},
Zikun Lin$^{4}$, and Chichuan Jin$^{1,3,5}$\\
$^{1}$Key Laboratory of Space Astronomy and Technology, National Astronomical Observatories, Chinese Academy of Sciences, Beijing 100101, China\\
$^{2}$Centre for Extragalactic Astronomy, Department of Physics, Durham University, South Road, Durham, DH1 3LE, UK\\
$^{3}$School of Astronomy and Space Science, University of Chinese Academy of Sciences, Beijing 100049, China\\
$^{4}$Department of Astronomy, Xiamen University, Xiamen, Fujian 361005, People's Republic of China\\
$^{5}$Institute for Frontier in Astronomy and Astrophysics, Beijing Normal University, Beijing 102206, China
}

\date{Accepted XXX. Received YYY; in original form ZZZ}
\pubyear{2026}

\newcommand{\rs}{R_{\rm S}}
\newcommand{\rin}{R_{\rm in}}
\newcommand{\rout}{R_{\rm out}}
\newcommand{\ledd}{L_{\rm Edd}}
\newcommand{\mdedd}{\dot M_{\rm Edd}}
\newcommand{\mdotd}{\dot m}
\newcommand{\teff}{T_{\rm eff}}
\newcommand{\tcol}{T_{\rm col}}
\newcommand{\tc}{T_{\rm c}}
\newcommand{\etal}{\lambda_{\tau}}
\newcommand{\etau}{\lambda_{u}}
\newcommand{\msun}{M_\odot}

\begin{document}
\label{firstpage}
\pagerange{\pageref{firstpage}--\pageref{lastpage}}
\maketitle

\begin{abstract}
Tidal disruption events (TDEs) can reach super-Eddington luminosities
near their peak. While their optical/UV emission may arise from multiple
processes, X-rays are more likely to be produced by the innermost
accretion flow close to the black hole, analogous to super-Eddington flows in
stellar-mass X-ray binaries (XRBs) and active galactic nuclei (AGNs).
Super-Eddington AGNs typically show disc-dominated spectra, with weak and steep coronae, while super-Eddington XRBs show a wider variety of spectral shapes. Here we explore disc-corona coupling in TDEs by developing a super-Eddington ($\dot{m}>1$: slim) disc model in which magnetic reconnection transfers some of the accretion power from the disc into a hot Comptonizing corona. We test the model on the well-monitored X-ray TDE AT2021ehb, and can 
broadly fit the brightest X-ray spectra with a black-hole mass 
$M_{\rm BH}\sim10^6$--$10^7\,M_\odot$ at $\dot m\sim 1$, but need variable magnetic coupling to the disc to match the large observed fast changes in coronal contribution. AGN at similarly high luminosity do not show such variable coronae, but the transient XRB outbursts do, suggesting that it may be a common feature of rapidly changing, non-steady accretion flows. As the TDE declines to $\dot{m}\sim 0.01$, the source hardens, but never goes below 
$\Gamma\sim 2$, unlike both AGN and XRB at these luminosities. Thus the X-ray spectral evolution in this TDE does not resemble either XRB or AGN in detail, likely pointing to the very different, highly variable nature of the accretion flow in a tidal disruption event.
\end{abstract}

\begin{keywords}
accretion, accretion discs -- black hole physics -- transients: tidal disruption events -- X-rays: galaxies
\end{keywords}

\section{Introduction}

When a star passes within the tidal radius of a massive black hole, it
can be partially or fully disrupted and a fraction of the stellar debris
can return to the black hole on highly eccentric orbits
\citep{Hills1975,Rees1988,Phinney1989,Evans1989,Guillochon2013}.
The resulting tidal disruption event (TDE) offers a rare opportunity to follow
the assembly of an accretion flow from its earliest stages. Because the fallback
rate can be close to, or even exceed, the Eddington rate, TDEs probe the regime
where radiation pressure and advection may be important
\citep{Strubbe2009,Lodato2011}. Winds, circularization and disc-corona coupling
can also shape the observed emission
\citep{Komossa2015,Komossa2017Tidal18,Auchettl2017,Gezari2021,Bu2022Radiative13,Bu2023Radiative4,Qiao2025Early12}.

The X-ray spectra of TDEs are especially valuable because they arise from the
innermost accretion flow. Many events are observed as very soft X-ray sources,
with temperatures of a few tens to hundreds of eV, while a subset develops a
harder X-ray component as the event evolves. Such hardening is often described
phenomenologically with a blackbody plus a power law. Although useful, this
approach does not directly connect the observed spectral shape to the underlying
disc structure, the coronal optical depth, or the fraction of accretion power
transported into the corona.

Disc-corona models provide a physical framework for this problem.
Energy-balance models couple a hot Comptonizing plasma to the soft photons
produced or reprocessed by an optically thick disc
\citep{Haardt1991,Haardt1993}. The hot phase may form a smooth layer above the
disc, or it may be concentrated in compact or patchy regions; thermal and
hybrid electron distributions can also produce different high-energy spectral
shapes \citep{Gierlinski1999,Done2007}. In magnetic models, buoyancy and
reconnection transport energy from the disc into the corona
\citep{Galeev1979,DiMatteo1998}. Detailed calculations then connect magnetic
heating, Compton cooling and the underlying disc structure
\citep{Liu2002,Liu2003,Cao2009,Liu2012,Cheng2020}. In TDEs, the accretion rate
can change by orders of magnitude over a few years. The same disc-corona
coupling can therefore produce strong spectral evolution. Recent work has
applied this idea to TDEs and related nuclear transients
\citep{Dai2018,Mageshwaran2023Disk5,Chen2026The2,Xu2025}. The changing X-ray
spectra can then be used to trace coronal formation and decay.

In this paper we extend the magnetic picture with a slim disc-corona model for
near- and super-Eddington accretion in TDEs. The calculation assumes a smooth
hot corona coupled locally to the entire disc. It follows the radial disc structure,
applies a colour correction to the disc radiation, and iterates the effective
coronal optical depth and soft-photon energy density to satisfy the adopted
energy balance. Compared with our previous TDE disc-corona calculation
\citep{Chen2026The2}, it solves the gas pressure, radiation pressure, magnetic
pressure, advection and coronal cooling in one coupled set of equations. This
gives smoother radial structures and more continuous spectra across different
pressure regimes. The detailed prescription is described in
Section~\ref{sec:model}.

We then apply the model to AT2021ehb. This nearby TDE, at $z=0.018$, showed
delayed X-ray brightening, a gradual soft-to-hard transition and a rapid return
to a softer, fainter state \citep{Yao2022AT2021ehb}. The bright hard interval contained a
strong hard X-ray component and was interpreted as evidence for the formation
of a magnetically dominated corona. Later XMM-Newton spectra also show a
thermal disc together with a harder tail and possible additional spectral
structure \citep{Xiang2024AT2021ehb}. AT2021ehb is therefore a useful X-ray
test of whether one physical disc-corona model can follow the overall evolution
from disc-dominated to strongly Comptonized spectra. We use it as an
application of the model, rather than attempting a complete multiwavelength
description of the event.

The paper is organized as follows. In Section~\ref{sec:model} we describe the
slim disc-corona model. Section~\ref{sec:model_results} presents representative
spectra and radial structures. Section~\ref{sec:at2021ehb} applies the model to
the Swift/XRT and XMM-Newton spectra of AT2021ehb. We discuss the implications
and limitations in Section~\ref{sec:discussion}, and summarize the main results
in Section~\ref{sec:conclusions}.

\section{Slim disc-corona model}
\label{sec:model}

\subsection{Physical picture and basic scalings}

We construct a radial, steady slim disc-corona model for accretion flows in TDEs.
The model follows the magnetic-reconnection disc-corona framework originally
developed for accretion discs around black holes
\citep{Haardt1993,DiMatteo1998,Liu2002,Liu2003,Cao2009,Liu2012,Liu2016,Cheng2020,Liu2024Hybrid}
and extends it to near- and super-Eddington accretion and to the compact
circularization radii expected in TDEs. In this picture, we assume that all the
mass accretes through the optically thick disc, with angular momentum
transported by a magnetic turbulent dynamo. Magnetic buoyancy transports the
fields upward, where reconnection heats the upper disc atmosphere and forms a static corona. This is slightly different to the accreting corona models of e.g. \citet{QiaoLiu2013,QiaoLiu2018}, where the mass in the corona can also move radially. 

The key quantity determining the spectrum is the fraction of the total heating
power dissipated in the corona, $f_{\rm c}$, rather than in the optically thick
disc. Older models \citep{Haardt1993,Svensson1994} used $f_{\rm c}$ directly,
usually assuming that it was constant with radius. In newer models, because the
heating ultimately derives from the magnetic field, the more fundamental
parameter is the ratio of disc pressure to magnetic pressure. Coronal heating is
balanced by Compton cooling, so specifying the magnetic heating allows the
coronal density and temperature to be determined. The model then obtains
$f_{\rm c}$ self-consistently as a function of radius when the pressure ratio
is held constant.

We follow this approach, described in \citet{Chen2026The2}, but change an
important part of the structural treatment. In the earlier calculation,
gas-pressure-dominated and
radiation-pressure-dominated solutions were handled separately. Here the gas
pressure, radiation pressure, magnetic pressure, advection and coronal cooling
are solved together within one coupled set of slim disc-corona equations. The
radial solution can therefore vary continuously between pressure regimes and
magnetic-field strengths, rather than switching between separate prescribed
branches.

For a black hole of mass $M_{\rm BH}$, the Schwarzschild radius and Eddington
luminosity are
\begin{equation}
  \rs = \frac{2GM_{\rm BH}}{c^2},
  \qquad
  \ledd = \frac{4\pi GM_{\rm BH} m_{\rm p} c}{\sigma_{\rm T}}.
\end{equation}
The dimensionless accretion rate is defined as
\begin{equation}
  \mdotd = \frac{\dot M}{\mdedd}, \qquad
  \mdedd = \frac{\ledd}{\eta c^2},
\end{equation}
where $\eta$ is a fixed radiative-efficiency parameter used to set the
Eddington accretion-rate scale. We use a pseudo-Newtonian radial structure with
inner edge $\rin=3\rs$ for the fiducial non-spinning case, and adopt
$\eta=1/16$ consistently with this choice. This follows the standard
slim-disc picture for near- and super-Eddington flows
\citep{Abramowicz1988,Watarai2000,Kitaki2018}.

\subsection{Local slim-disc equations}

At each radius $R$, the disc part of the calculation solves for the mid-plane
density $\rho_{\rm d}$, disc temperature $T_{\rm d}$ and scaleheight $H$ under
vertical hydrostatic equilibrium. The local viscous energy release is
\begin{equation}
  Q_{\rm vis}^{+}(R) =
  \frac{3GM_{\rm BH}\dot M}{8\pi R^3}
  \left[1-\left(\frac{\rin}{R}\right)^{1/2}\right],
  \label{eq:qvis}
\end{equation}
where the bracket is denoted by $\phi(R)$ below. Radial advection is included
through
\begin{equation}
  Q_{\rm adv}^{-} =
  \frac{\dot M}{4\pi R^2}\frac{p}{\rho_{\rm d}}\xi,
  \label{eq:qadv}
\end{equation}
where we set $\xi=1.5$, an order-unity factor describing the radial entropy
gradient.
The remaining accretion power is divided between the optically thick disc and
the corona. If $f_{\rm c}$ is the fraction transported into the corona, the
disc radiative cooling satisfies
\begin{equation}
  \left(Q_{\rm vis}^{+}-Q_{\rm adv}^{-}\right)(1-f_{\rm c})
  = Q_{\rm rad}^{-},
  \label{eq:disc_energy}
\end{equation}
with
\begin{equation}
  Q_{\rm rad}^{-}=\frac{4\sigma T_{\rm d}^{4}}{3\kappa\Sigma},
  \qquad
  \Sigma=2\rho_{\rm d}H.
  \label{eq:qrad}
\end{equation}
Here $\kappa=0.4\,{\rm cm^2\,g^{-1}}$ is the electron-scattering opacity used
for the optically thick disc.
The angular momentum equation is written in the usual alpha-viscosity form,
\begin{equation}
  \dot M(1-f_{\rm c})\Omega_{\rm K}
  \left[1-\left(\frac{\rin}{R}\right)^{1/2}\right]
  =4\pi H\alpha p,
  \label{eq:angular}
\end{equation}
where $\Omega_{\rm K}$ is the Keplerian angular velocity and
$\alpha$ is the viscosity parameter. The total pressure is
\begin{equation}
  p=p_{\rm rad}+p_{\rm gas}+p_{\rm B},
  \label{eq:pressure_sum}
\end{equation}
with
\begin{equation}
  p_{\rm rad}=\frac{aT_{\rm d}^{4}}{3},
  \quad
  p_{\rm gas}=\frac{\rho_{\rm d}kT_{\rm d}}{\mu m_{\rm H}},
  \quad
  p_{\rm B}=\frac{B^2}{8\pi}.
  \label{eq:pressure_components}
\end{equation}
In the gas-pressure term we use $\mu=0.5$ for a fully ionized hydrogen-rich
plasma.
The magnetic field strength is specified by
\begin{equation}
  \beta = \frac{p_{\rm gas}+p_{\rm rad}}{p_{\rm B}},
  \label{eq:beta_def}
\end{equation}
so that smaller $\beta$ means a stronger magnetic field. We further impose
$p/\rho_{\rm d}=\Omega_{\rm K}^{2}H^2$. The disc structure and the coronal
fraction $f_{\rm c}$ are solved together with the coronal energy balance below.

\subsection{Magnetic-reconnection corona}

The coronal fraction is not prescribed by hand. It is set by the upward
magnetic energy flux released by reconnection,
\begin{equation}
  Q_{\rm cor}^{-}=\frac{B^2}{4\pi}V_{\rm A},
  \qquad
  V_{\rm A}=\frac{B}{(4\pi\mu m_{\rm H}n_{\rm c})^{1/2}},
  \label{eq:qcor}
\end{equation}
where $n_{\rm c}$ is the coronal electron density and $V_{\rm A}$ is the
Alfv\'en speed. We identify the fraction of local accretion power dissipated in
the corona as
\begin{equation}
  f_{\rm c} =
  \frac{Q_{\rm cor}^{-}}{Q_{\rm vis}^{+}-Q_{\rm adv}^{-}}.
  \label{eq:fc_def}
\end{equation}
The coronal temperature $T_{\rm e}$ and density $n_{\rm c}$ are then obtained
from two balance equations. First, magnetic heating balances inverse Compton
cooling,
\begin{equation}
  \frac{B^2}{4\pi}V_{\rm A}
  \simeq
  \frac{4kT_{\rm e}}{m_{\rm e}c^2}\tau^{*}cU_{\rm rad}.
  \label{eq:compton_balance}
\end{equation}
Second, electron thermal conduction into the disc is balanced by the enthalpy
flux of evaporating material,
\begin{equation}
  \frac{k_0T_{\rm e}^{7/2}}{l_{\rm c}}
  \simeq
  \frac{\gamma}{\gamma-1}n_{\rm c}kT_{\rm e}
  \left(\frac{kT_{\rm e}}{m_{\rm H}}\right)^{1/2}.
  \label{eq:conduction_balance}
\end{equation}
Here $\gamma=5/3$ is the adopted adiabatic index and
$k_0=10^{-6}$ in cgs units is the Spitzer conductivity coefficient. We
take the characteristic coronal length to be $l_{\rm c}=H$, assuming that the
disc scaleheight sets the scale of the magnetic loops. Thus, the effective
scattering depth is
\begin{equation}
  \tau^{*}=\etal n_{\rm c}\sigma_{\rm T}l_{\rm c}.
  \label{eq:taustar}
\end{equation}

The radiation energy density entering the Compton cooling includes the disc
radiation field and reprocessed coronal photons,
\begin{equation}
  U_{\rm rad}=\frac{2}{c}Q_{\rm rad}^{-}
  +0.4\,\etau\,\frac{B^2}{8\pi}.
  \label{eq:urad}
\end{equation}
At each radius, we solve equations~(\ref{eq:disc_energy}),
(\ref{eq:angular}), (\ref{eq:compton_balance}), and
(\ref{eq:conduction_balance}) simultaneously for $\rho_{\rm d}$,
$T_{\rm d}$, $n_{\rm c}$, and $T_{\rm e}$, using the auxiliary relations
defined above.

\subsection{Global consistency, colour correction and spectra}

The local equations above contain two correction factors. The first,
$\etal$, corrects the effective optical depth used in Compton cooling. The
second, $\etau$, corrects the local soft-photon energy density. They are
introduced because the local slab approximation does not by itself guarantee
global energy consistency once the spectrum is integrated over radius. We start
the iteration from $\etal=\etau=1$.

For each trial radial solution we compute the gravitational power and advected
power,
\begin{equation}
  \begin{aligned}
  L_{\rm G} &=
  \int_{\rin}^{\rout}2\pi R
  \frac{3GM_{\rm BH}\dot M\phi(R)}{8\pi R^3}\,{\rm d}R,\\
  L_{\rm adv} &= \int_{\rin}^{\rout}2\pi RQ_{\rm adv}^{-}\,{\rm d}R.
  \end{aligned}
  \label{eq:lg_ladv}
\end{equation}
The intrinsic soft luminosity is
\begin{equation}
  L_{\rm soft}=\int_{\rin}^{\rout}2\pi R\sigma T_{\rm R}^{4}\,{\rm d}R,
  \label{eq:lsoft}
\end{equation}
where $T_{\rm R}$ is the local effective temperature of the soft radiation
field. The upward and downward Comptonized luminosities,
$L_{\rm up}$ and $L_{\rm down}$, are then used to adjust the two correction
factors. For the upward luminosity, energy conservation requires
\[
  \frac{L_{\rm up}+0.2L_{\rm down}}{L_{\rm G}-L_{\rm adv}}\simeq 1,
  \qquad
  \lambda_{\tau,n+1}
  =\frac{L_{\rm up}+0.2L_{\rm down}}{L_{\rm G}-L_{\rm adv}}\lambda_{\tau,n}.
\]
For the downward luminosity, the absorbed part of the downward coronal emission
is compared with the intrinsic soft luminosity,
\[
  \frac{0.8L_{\rm down}}{L_{\rm soft}}\simeq \lambda'_{u,n},
  \qquad
  \lambda_{u,n+1}
  =\frac{0.8L_{\rm down}/L_{\rm soft}}{\lambda'_{u,n}}\lambda_{u,n}.
\]
Here we adopt an energy-independent disc albedo of 0.2. Thus 20\% of
$L_{\rm down}$ is treated as reflected or scattered radiation, while the
remaining 80\% is absorbed by the disc surface and reprocessed into the
soft seed-photon field. In practice, we update $\etal$ and $\etau$ in the
iterative calculation until the two conditions above are satisfied. This
procedure modifies the local soft seed-photon density and the effective
scattering depth used in the Monte Carlo calculation of the disc-corona
structure and spectrum. Detailed descriptions of this iterative procedure can
be found in \citet{Liu2003}, \citet{Liu2016} and \citet{Cheng2020}.

For the representative sequence with $M_{\rm BH}=10^6\,M_\odot$ and
$\beta=50$, the converged values of $(\lambda_{\tau},\lambda_{u})$ are
$(1.47,0.27)$, $(1.16,1.78)$, and $(1.07,7.51)$ for $\dot m=0.03$, $0.3$,
and $10$, respectively. The solutions converge within 7--8 global iterations
using the adopted 5 per cent tolerance. Thus, $\lambda_{\tau}$ remains relatively
close to unity, whereas the soft-photon correction $\lambda_{u}$ varies more
strongly and becomes substantially larger than unity in the high-$\dot m$
solution. These quantities are global energy-balance corrections rather than
additional free parameters of the spectral fits.

The effective disc temperature follows from $\sigma \teff^4=Q_{\rm rad}^{-}$.
Because electron scattering hardens the emergent disc atmosphere, we use a
colour correction factor $f_{\rm col}$ and set
\begin{equation}
  \tcol=f_{\rm col}\teff.
\end{equation}
We use the temperature-dependent colour correction prescription of
\citet{Done2012},
\begin{equation}
  f_{\rm col} =
  \begin{cases}
  1, & \teff < 3\times10^4\,{\rm K},\\
  \left(\teff/3\times10^4\,{\rm K}\right)^{0.82},
  & 3\times10^4\,{\rm K}\le \teff < 10^5\,{\rm K},\\
  \left(72\,{\rm keV}/k\teff\right)^{1/9},
  & \teff \ge 10^5\,{\rm K}.
  \end{cases}
  \label{eq:fcol}
\end{equation}
At low temperatures the disc spectrum is close to a blackbody. At higher
temperatures electron scattering becomes increasingly important, so the
emergent colour temperature is harder than the effective temperature.
At each radius, the emergent emission is the sum of the diluted
colour-corrected thermal component and the Comptonized coronal component. For
the latter, seed photons with colour temperature $\tcol$ are injected into a hot
slab specified by $T_{\rm e}$ and $\tau^{*}$, and inverse Compton scattering is
followed with Monte Carlo simulations \citep{Pozdnyakov1977,Pozdnyakov1983}. This gives the upward escaping spectrum
and the downward spectrum that irradiates the disc, rather than imposing a
phenomenological power-law tail. The total spectrum is then obtained by
integrating the local emergent intensity over the whole disc.
The emergent spectra are evaluated on a logarithmically spaced energy grid
from $10^{-3}$ to $10^{3}\,\mathrm{keV}$ using 500 bins, corresponding to
approximately $0.012$ dex per bin. This range covers the optical/UV disc
emission as well as the X-ray coronal continuum.

\section{Model spectra and radial structure}
\label{sec:model_results}

We explore the dependence of the model spectra and radial structure
on three parameters: the black-hole mass, $M_{\rm BH}$, the dimensionless
accretion rate, $\mdotd$, and the magnetic pressure parameter, $\beta$.
Throughout the calculations presented here, we fix the viscosity parameter
at $\alpha=0.3$. We vary one of the three parameters at a time to illustrate
its effect before applying the model to data.

\begingroup
For the TDE calculations presented here, we adopt a solar-type disrupted
star, with $M_*=\msun$ and $R_*=R_\odot$. The tidal radius and
circularization radius are
\begin{equation}
  R_{\rm t}=R_{*}\left(\frac{M_{\rm BH}}{M_{*}}\right)^{1/3},
  \qquad
  R_{\rm c}\simeq2R_{\rm t},
  \label{eq:rt_rc}
\end{equation}
and we set
\begin{equation}
  \rout=R_{\rm c}.
  \label{eq:rout_rc}
\end{equation}
The outer radius is therefore not an independent fitting parameter in
the TDE table model used here.
\par\endgroup

The spectra have two physically distinct components. The optical/UV-to-EUV
emission is mainly set by the multicolour blackbody radiation from the optically
thick disc, while the soft X-ray tail is produced by inverse Compton scattering
of disc photons in the magnetically heated corona. The disc component is
therefore mostly controlled by the temperature scale and emitting area, whereas
the X-ray spectral shape is sensitive to the coronal heating fraction,
temperature, and effective scattering depth. In the following subsections, we
show that $\mdotd$ mainly changes the luminosity scale and the strength of
Compton cooling, $\beta$ most directly regulates the fraction of accretion power
released in the corona, and $M_{\rm BH}$ shifts the characteristic disc
temperature and luminosity scale.

\subsection{The effect of mass accretion rate}

\begin{figure}
  \centering
  \includegraphics[width=\columnwidth]{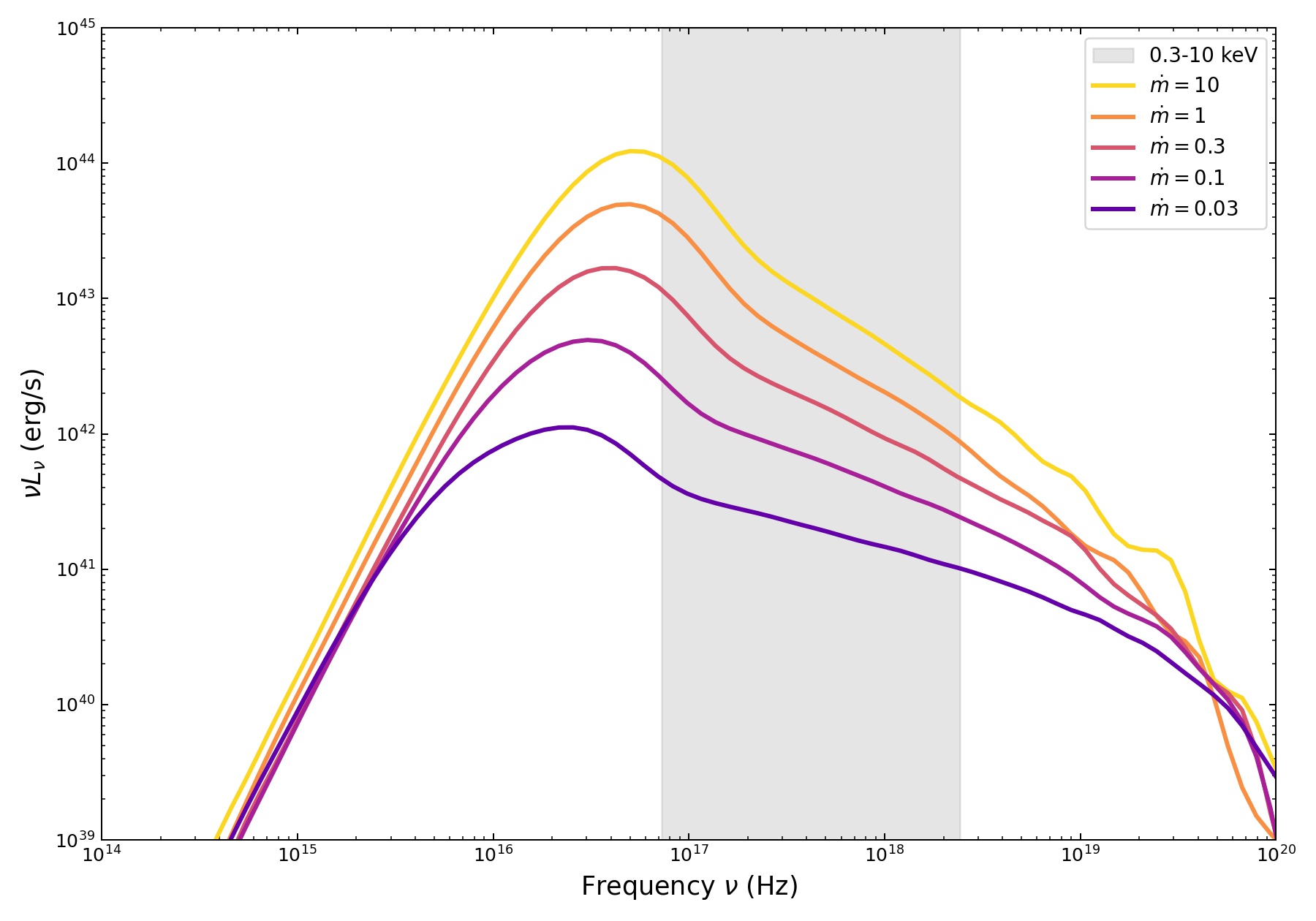}
  \caption{Emergent spectra for $M_{\rm BH}=10^6\,\msun$ and $\beta=50$ at
  different accretion rates. As $\mdotd$ declines, the bolometric luminosity
  decreases and the relative contribution of the coronal tail changes across
  the 0.3--10 keV band.}
  \label{fig:model_mdot_spectrum_1e6}
\end{figure}

Fig.~\ref{fig:model_mdot_spectrum_1e6} shows the spectra for
$M_{\rm BH}=10^6\,\msun$ and $\beta=50$ with
$\mdotd=10,\,1,\,0.3,\,0.1,$ and $0.03$. As expected, increasing $\mdotd$
raises the overall luminosity because more gravitational power is supplied to
the disc--corona system, and it also raises the disc temperature.
The spectral change in the X-ray band, however, is not simply a
renormalisation of the coronal component. At high $\mdotd$, the larger soft-photon
field cools the corona more efficiently and the spectrum is relatively
disc-dominated. As $\mdotd$ decreases, the soft-photon cooling weakens and a
larger fraction of the emergent luminosity can appear in the Comptonized tail.
The sequence therefore naturally produces a smooth soft-to-hard evolution during a
decline in accretion rate, but the hardest spectra cannot go below $\Gamma\sim 2$ due to the lower limit on seed photons from reprocessing of the irradiating corona flux.

\begin{figure*}
  \centering
  \includegraphics[width=0.86\textwidth]{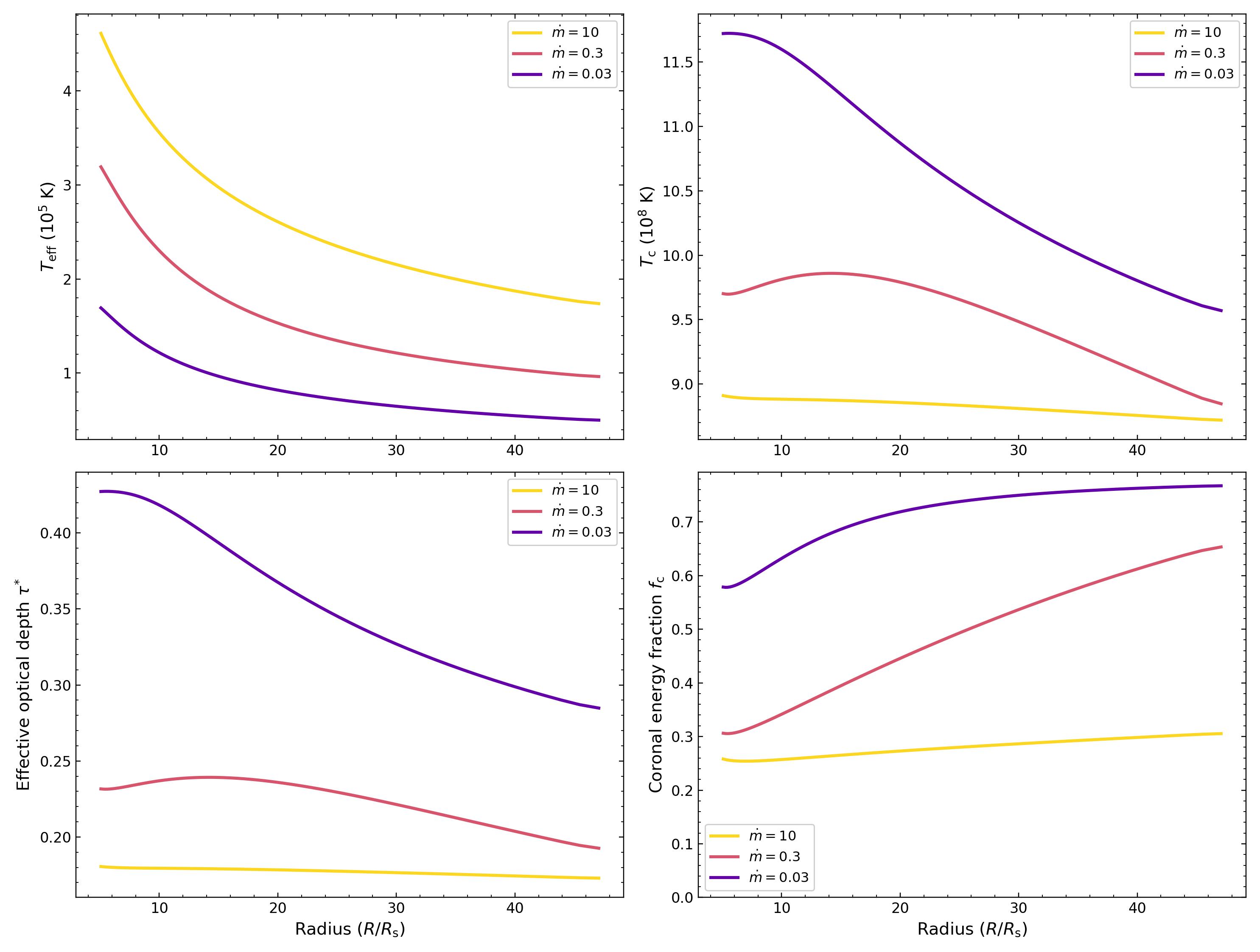}
  \caption{Radial structures corresponding to the model in
  Fig.~\ref{fig:model_mdot_spectrum_1e6}. The panels show $\teff$, $\tc$,
  the effective optical depth $\tau^{*}$, and the coronal energy fraction
  $f_{\rm c}$.}
  \label{fig:model_mdot_structure_1e6}
\end{figure*}

The corresponding radial structures are shown in
Fig.~\ref{fig:model_mdot_structure_1e6}. The effective temperature of the disc
decreases with decreasing $\mdotd$ at all radii, reflecting the lower local
dissipation rate. The coronal quantities respond in the opposite direction in
the inner disc: the lower-$\mdotd$ solutions have a hotter corona, a larger
effective optical depth $\tau^{*}$, and a larger coronal energy fraction
$f_{\rm c}$. This behaviour is a consequence of the self-consistent coupling
between magnetic heating and Compton cooling. When the disc radiation field is
weaker, the same magnetic-reconnection channel can maintain a hotter corona and
the reprocessed radiation becomes a more important part of the local soft-photon
budget. The iterative $\etal$ and $\etau$ corrections are important here
because they enforce the global energy balance described in
Section~\ref{sec:model}.

\begin{figure}
  \centering
  \includegraphics[width=\columnwidth]{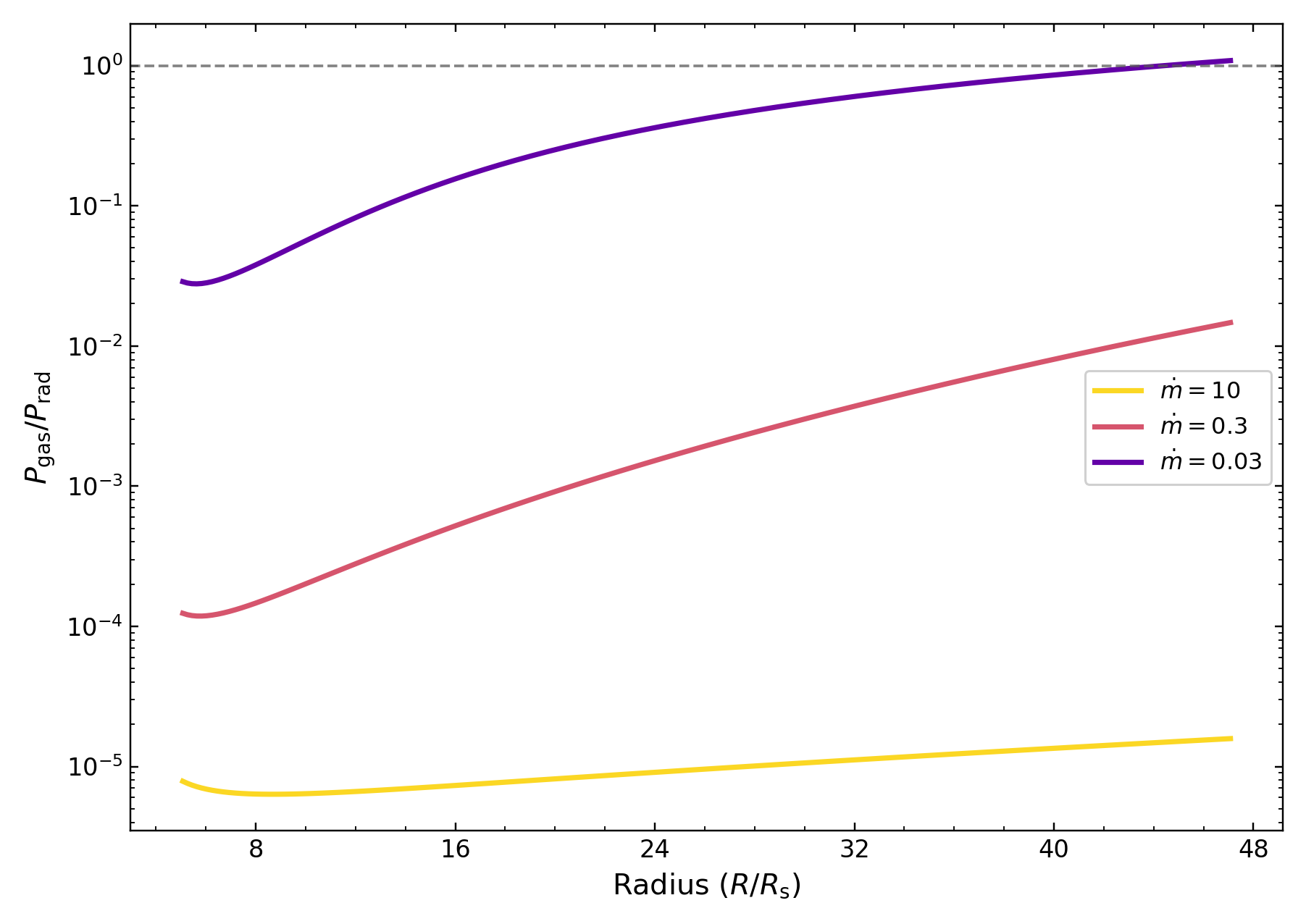}
  \caption{Ratio of gas pressure to radiation pressure for the same
  $M_{\rm BH}=10^6\,\msun$ and $\beta=50$ models shown in
  Fig.~\ref{fig:model_mdot_structure_1e6}. The dashed line marks
  $P_{\rm gas}=P_{\rm rad}$. As the accretion rate decreases, the outer disc
  moves closer to the gas-pressure-dominated regime.}
  \label{fig:model_mdot_pgas_prad_1e6}
\end{figure}

The pressure ratio in Fig.~\ref{fig:model_mdot_pgas_prad_1e6} provides a
compact diagnostic of the pressure regime occupied by these solutions. The
high-$\mdotd$ model remains strongly radiation-pressure-dominated over the
computed radial range, while the lower-$\mdotd$ solutions have progressively
larger $P_{\rm gas}/P_{\rm rad}$. This trend helps explain why the radial
coronal structure changes as the accretion rate declines: the disc density,
scaleheight and magnetic pressure are all tied to the changing balance between
radiation and gas pressure.

\subsection{The effect of magnetic-field strength}

The magnetic parameter $\beta$ is defined as the ratio of the gas-plus-radiation
pressure to the magnetic pressure. A smaller $\beta$ therefore means a stronger
magnetic field. Since the coronal heating rate scales with the magnetic energy
flux, a smaller $\beta$ increases the amount of accretion power transported
from the disc into the corona. This makes $\beta$ the most direct control
parameter for the X-ray strength in the present model.

\begin{figure}
  \centering
  \includegraphics[width=\columnwidth]{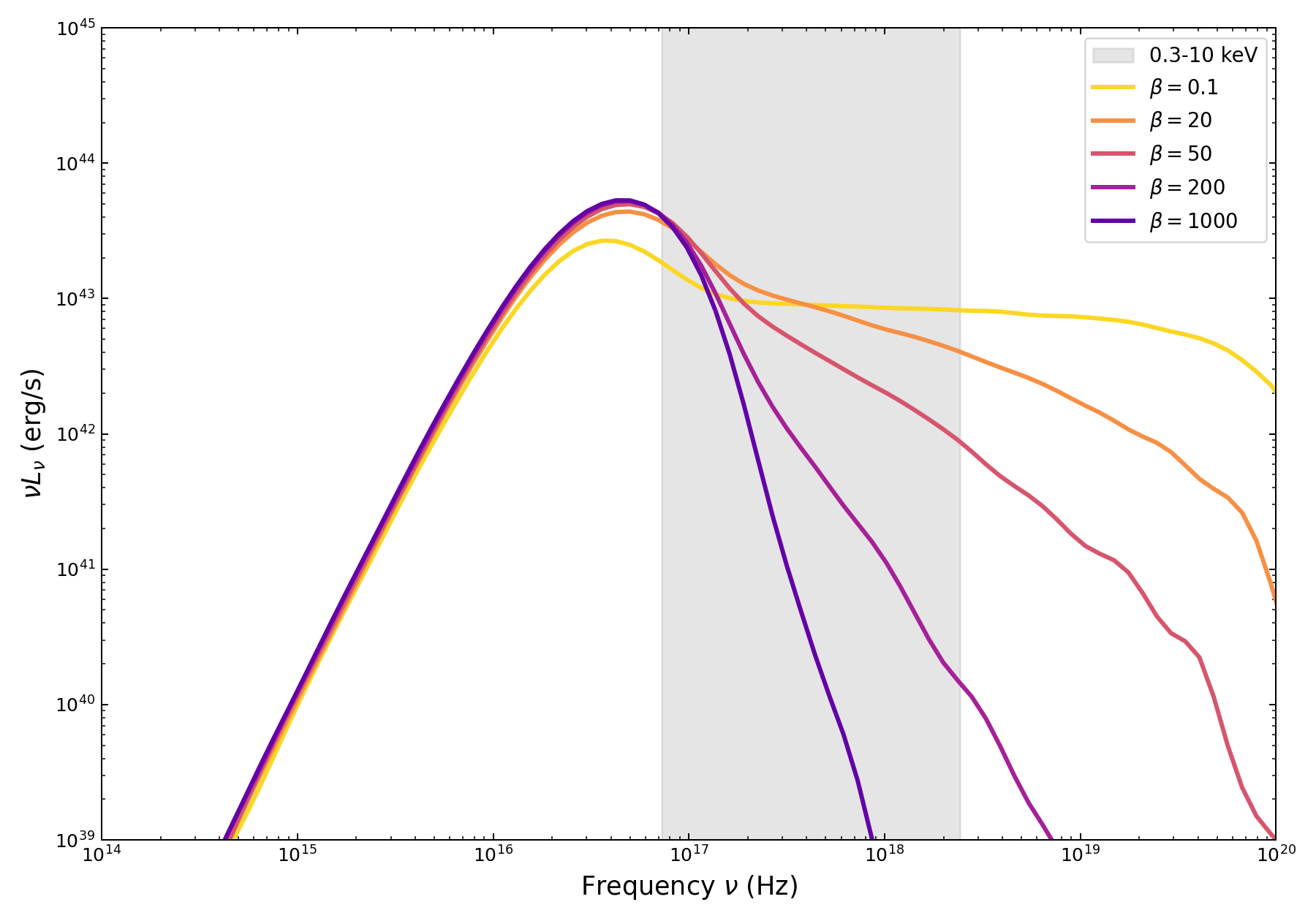}
  \caption{Effect of the magnetic parameter $\beta$ on the emergent spectrum
  for $M_{\rm BH}=10^6\,\msun$ and $\mdotd=1$. Smaller $\beta$ corresponds to a
  stronger magnetic field and a more prominent high-energy tail.}
  \label{fig:model_beta_spectrum_1e6}
\end{figure}

Fig.~\ref{fig:model_beta_spectrum_1e6} demonstrates this effect for
$M_{\rm BH}=10^6\,\msun$ and $\mdotd=1$. The disc peak is only weakly affected
by changing $\beta$, because the characteristic disc temperature is primarily
set by $M_{\rm BH}$ and $\mdotd$. The high-energy tail, in contrast, changes
dramatically. We show a broad sequence from the extreme strong-field case
$\beta=0.1$ to the weak-field case $\beta=1000$. For small $\beta$, the
stronger magnetic field powers a prominent Comptonized component extending
across the 0.3--10 keV band. For
$\beta=1000$, the magnetic field is weak, only a small fraction of the
gravitational energy is dissipated in the corona, and the spectrum approaches a
disc-dominated shape with a much weaker X-ray tail.

\begin{figure*}
  \centering
  \includegraphics[width=0.86\textwidth]{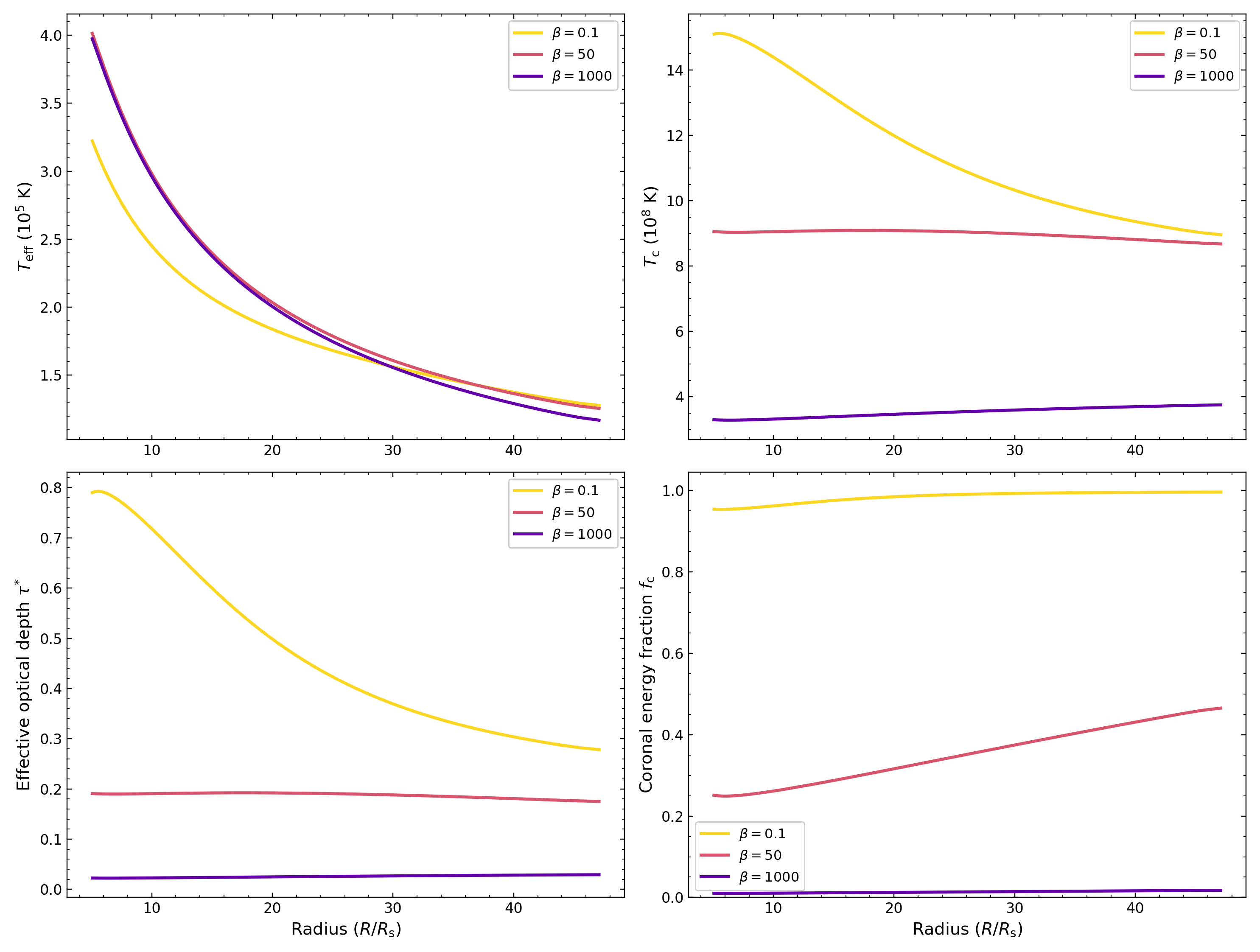}
  \caption{Radial structures for three representative $\beta$ values in the
  spectral sequence shown in Fig.~\ref{fig:model_beta_spectrum_1e6}. The
  coronal temperature and coronal energy fraction respond strongly to the
  magnetic-field strength, whereas the disc effective temperature is less
  sensitive.}
  \label{fig:model_beta_structure_1e6}
\end{figure*}

The radial structures in Fig.~\ref{fig:model_beta_structure_1e6} show why the
spectra are so sensitive to $\beta$. The disc effective temperature changes
only mildly among the three cases, but the coronal temperature, effective
optical depth, and energy fraction vary by large factors. A stronger magnetic
field drives a larger Alfv\'en energy flux into the corona, evaporates more
material into the hot phase, and increases both the Comptonizing optical depth
and the coronal luminosity fraction. Thus the model does not need to prescribe
an arbitrary coronal fraction: the hard X-ray strength follows from the local
magnetic pressure and the disc--corona energy balance. This is the same
physical reason why magnetic-reconnection corona models can generate a broad
range of optical-to-X-ray spectral shapes with a limited number of parameters.

\begin{figure}
  \centering
  \includegraphics[width=\columnwidth]{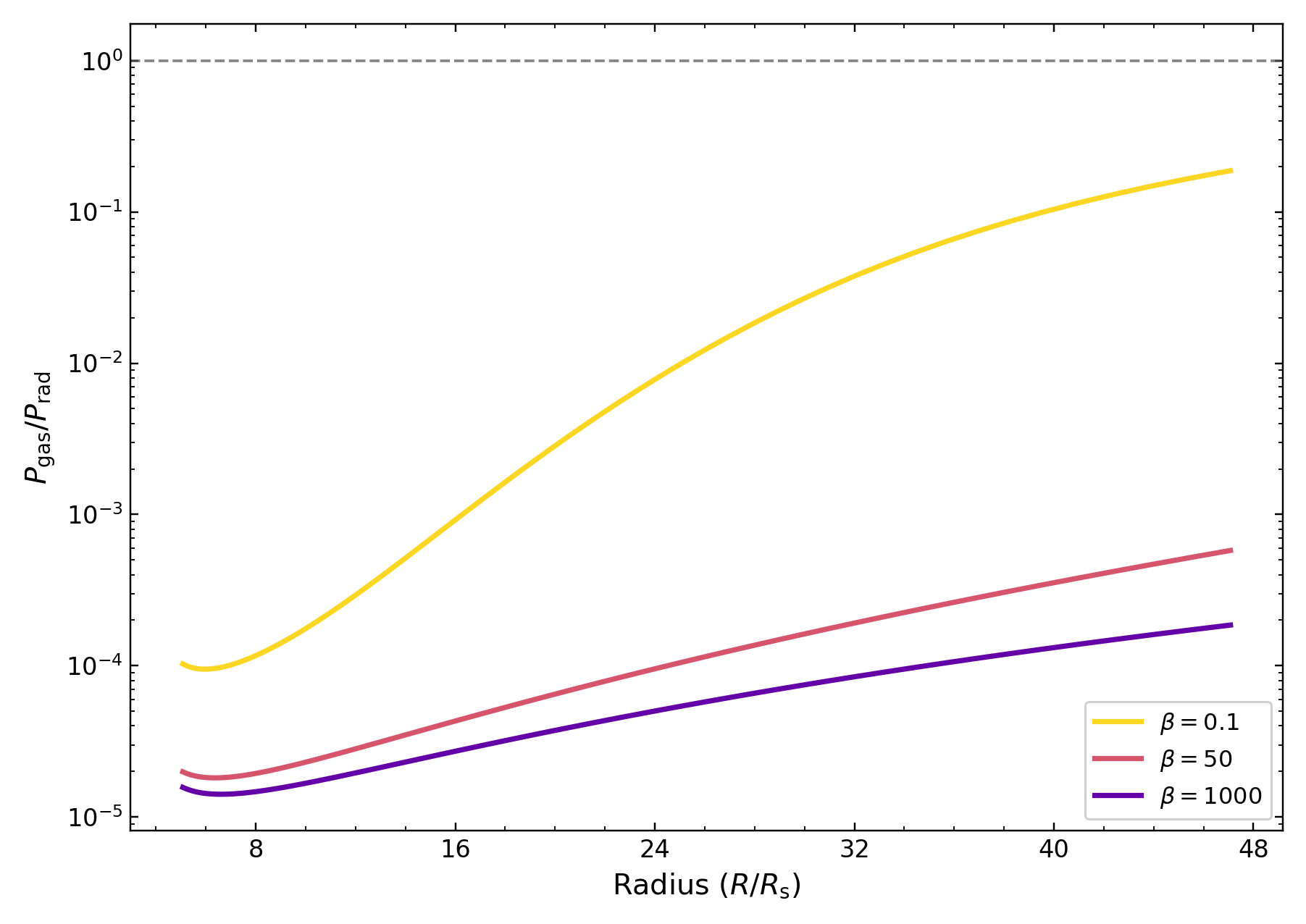}
  \caption{Ratio of gas pressure to radiation pressure for the representative
  $\beta$ sequence at $M_{\rm BH}=10^6\,\msun$ and $\mdotd=1$. The dashed line
  marks $P_{\rm gas}=P_{\rm rad}$. Stronger magnetic fields modify the local
  disc-corona solution and raise $P_{\rm gas}/P_{\rm rad}$, although the
  fiducial models remain radiation-pressure-dominated over most of the plotted
  radial range.}
  \label{fig:model_beta_pgas_prad_1e6}
\end{figure}

The pressure-ratio profiles in Fig.~\ref{fig:model_beta_pgas_prad_1e6} show
that the extreme strong-field solution also changes the underlying disc pressure
balance. This reinforces the interpretation that $\beta$ is not only a
phenomenological control of the X-ray tail, but enters through the local
magnetic, thermal and hydrostatic structure of the disc-corona system.

\subsection{The effect of black-hole mass}

\begin{figure}
  \centering
  \includegraphics[width=\columnwidth]{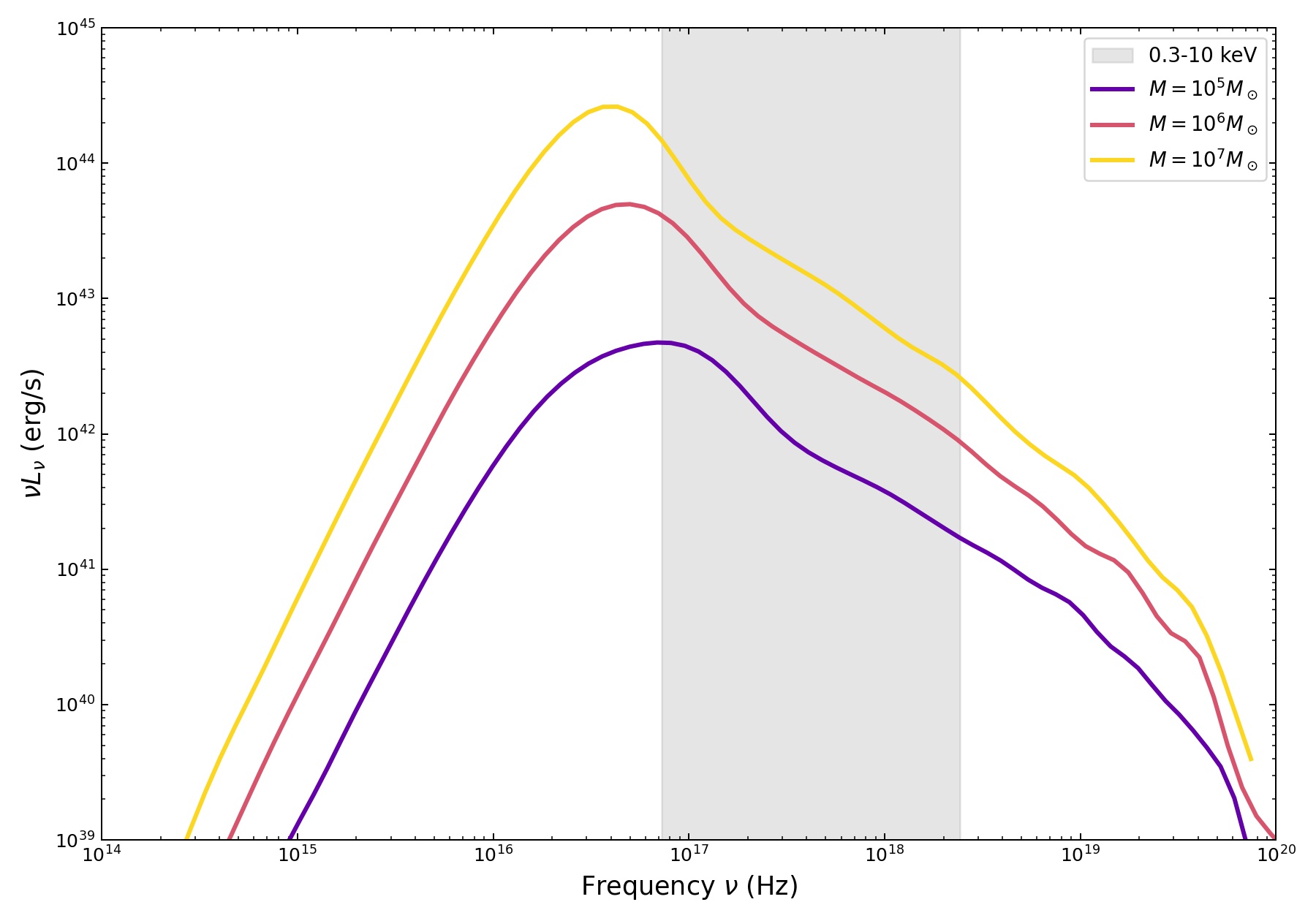}
  \caption{Mass dependence of the emergent spectrum at fixed $\mdotd=1$ and
  $\beta=50$. The thermal peak moves to lower frequency with increasing
  black-hole mass, while the absolute luminosity increases approximately with
  the Eddington scale.}
  \label{fig:model_mass_spectrum}
\end{figure}

Finally, Fig.~\ref{fig:model_mass_spectrum} compares models with
$M_{\rm BH}=10^5,\,10^6,$ and $10^7\,\msun$ at fixed $\mdotd=1$ and
$\beta=50$. At a fixed Eddington-scaled accretion rate, the bolometric
luminosity approximately scales with $L_{\rm Edd}\propto M_{\rm BH}$, while the
characteristic disc temperature decreases with increasing mass. Therefore the
larger-mass models are more luminous in absolute $\nu L_\nu$, but their thermal
peaks are shifted to lower characteristic frequencies. Conversely, lower-mass
black holes have hotter discs and can place a larger fraction of their thermal
emission closer to the soft X-ray band.

These changes in disc temperature and luminosity affect the part of the
spectrum sampled by the X-ray band. The spectrum also depends on the
accretion rate, $\mdotd$, and the magnetic pressure parameter, $\beta$.
Mass estimates from spectral fitting therefore depend on how well these
other parameters are constrained.

\section{Application to AT2021ehb}
\label{sec:data}
\label{sec:at2021ehb}

AT2021ehb (ZTF21aanxhjv) is a nearby, optically selected TDE at $z=0.018$,
corresponding to a luminosity distance of approximately 78 Mpc
\citep{Yao2022AT2021ehb,Xiang2024AT2021ehb}. Its delayed brightening, gradual
soft-to-hard transition, and factor of about ten X-ray decline within three
days provide a useful test of the model across strongly changing disc-corona
states \citep{Yao2022AT2021ehb}.

\subsection{X-ray data and long-term evolution}

We reduced the XMM-Newton EPIC-pn data using version 22.1.0 of the
XMM-Newton Science Analysis System (SAS; \citealt{Gabriel2004}). Calibrated
event lists were produced with \texttt{epproc}. We excluded intervals during
which the 10--12~keV background count rate exceeded 0.4 counts~s$^{-1}$, and
selected events using \texttt{PATTERN<=4} and \texttt{FLAG==0}. Source events
were extracted from a circular region of radius 30$^{\prime\prime}$, while
background events were extracted from a source-free circular region of the
same radius on the same CCD. Redistribution matrix files and ancillary
response files were generated with \texttt{rmfgen} and \texttt{arfgen},
respectively. We grouped the spectra with \texttt{specgroup}, adopting
\texttt{mincounts=25} and \texttt{oversample=3}.

We reprocessed the \textit{Swift}/XRT observations with \texttt{xrtpipeline}
to obtain cleaned event files and exposure maps. Source and background
spectra were extracted with \texttt{xselect}. The source region was a circle
of radius 30$^{\prime\prime}$, and the background region was a concentric
annulus with inner and outer radii of 60$^{\prime\prime}$ and
120$^{\prime\prime}$, respectively. The spectra were grouped with
\texttt{grppha} using minimum-count thresholds of 10 counts per bin for s1
and s9, 25 counts per bin for s2--s8, and 15 counts per bin for s10--s16.

\begin{figure*}
  \centering
  \includegraphics[width=0.80\textwidth]{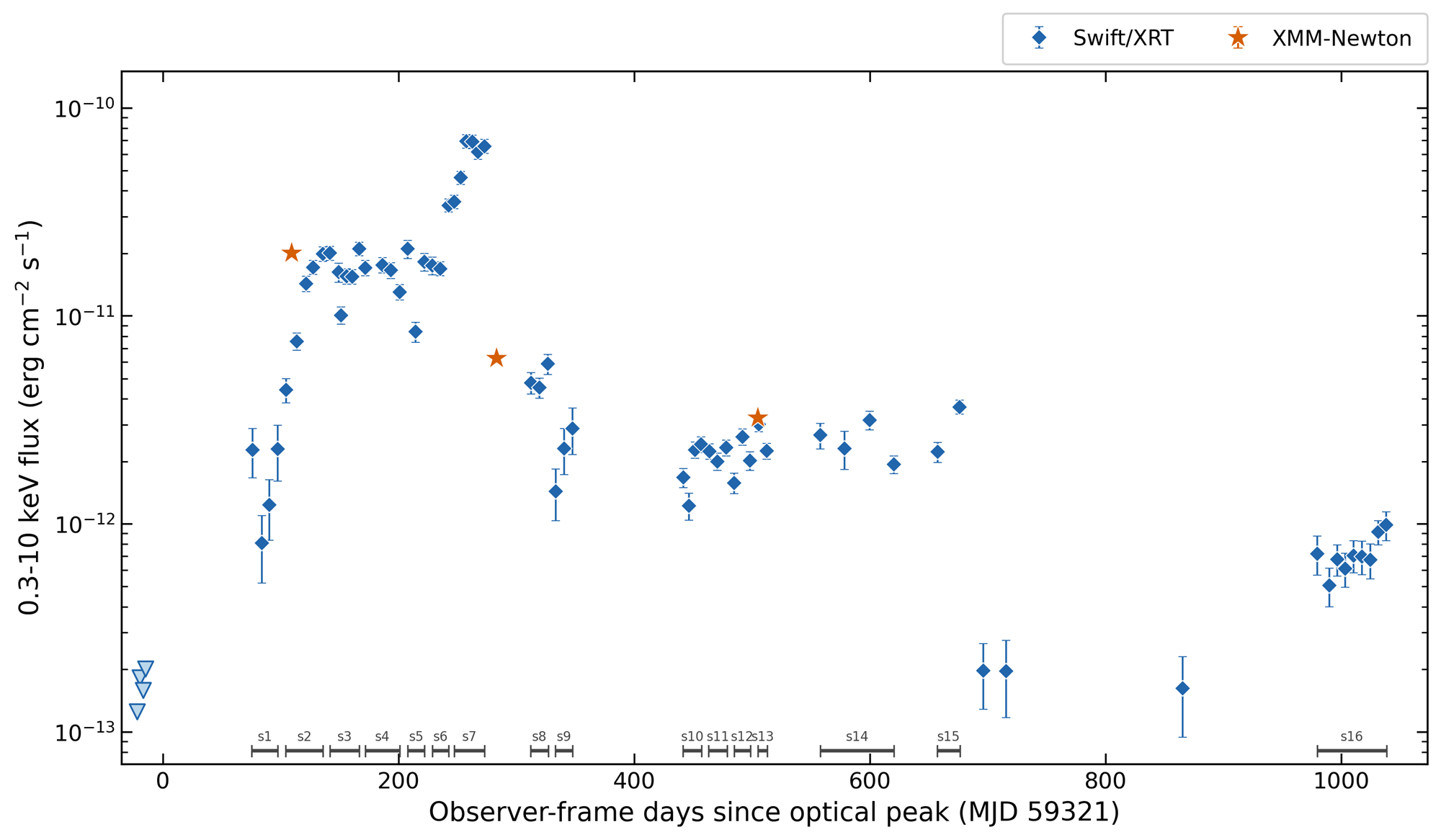}
  \caption{Long-term absorption-corrected 0.3--10 keV flux evolution of
  AT2021ehb. Blue diamonds show Swift/XRT observations and orange stars show
  XMM-Newton observations. Horizontal bars show the intervals used to
  construct the 16 stacked Swift spectra s1--s16, and downward triangles mark
  upper limits. The reference time is the optical peak at MJD 59321.}
  \label{fig:at2021ehb_lightcurve}
\end{figure*}

These Swift/XRT and XMM-Newton observations trace the X-ray evolution. The
long-term light curve is shown in
Fig.~\ref{fig:at2021ehb_lightcurve}. The plotted flux is the 0.3--10 keV flux
corrected only for Galactic neutral absorption. We do not apply a bolometric
correction because the disc peak can lie outside the observed band, making the
resulting extrapolation model dependent.

We also use XMM-Newton target-of-opportunity observations that provide
high signal-to-noise spectra. We choose three representative XMM-Newton EPIC-pn
spectra: X1 on 2021 August 4, X2
on 2022 January 25, and X3 on 2022 September 3 (see also
\citealt{Yao2022AT2021ehb}).

While individual Swift/XRT observations provide flux measurements, their
signal-to-noise is insufficient for spectral fitting. We follow
\citet{Yao2022AT2021ehb} for the first part of these data, co-adding
neighbouring Swift observations with similar flux and hardness. Extending this
grouping to later times gives 16 spectra, s1--s16, as shown in
Fig.~\ref{fig:at2021ehb_lightcurve}.
Three additional Swift observations between s15 and s16 had too few counts
for spectral fitting. We used these observations to derive count-rate upper
limits and, assuming a fixed spectral shape, indicative upper limits on
$\mdotd_{\rm XRT}$.\footnote{Observations 00014217065 and 00014217066 were
combined because they were separated by approximately 20 d and had consistent
count rates; observation 00014217067, obtained approximately 150 d later,
was treated separately. For each group, we calculated a one-sided 90 per cent
Poisson count-rate upper limit and converted it to an indicative
$\mdotd_{\rm XRT}$ limit by scaling relative to a nearby fitted spectral
state. This conversion assumes an unchanged spectral shape and extrapolates
below the table-model boundary at $\mdotd=0.01$.}
These do not provide independent spectral constraints on $\beta$.

\subsection{Spectral table model and fitting strategy}

The calculated spectra were tabulated as the additive XSPEC model
\texttt{slimdc.fits}. The table contains the black-hole mass, the dimensionless
mass accretion rate, and the magnetic parameter. We use the fixed
viscosity parameter and TDE outer-radius prescription specified in
Section~\ref{sec:model_results}. The parameter grid is
summarized in Table~\ref{tab:model_grid} and contains
$15\times21\times21=6615$ spectra.

\begin{table}
  \centering
  \caption{Parameter grid of the XSPEC table model \texttt{slimdc.fits}.}
  \label{tab:model_grid}
  \begin{tabular}{lccc}
    \toprule
    Parameter & Range & Step & Number \\
    \midrule
    $\log(M_{\rm BH}/M_\odot)$ & 4.0--7.5 & 0.25 & 15 \\
    $\log\mdotd$              & $-2.0$--3.0 & 0.25 & 21 \\
    $\log\beta$               & $-2.0$--3.0 & 0.25 & 21 \\
    \bottomrule
  \end{tabular}
\end{table}

We fit all spectra over 0.3--10 keV with
\texttt{TBabs*atable\{slimdc.fits\}}. The \texttt{TBabs} component uses the
ISM abundance table of \citet{Wilms2000}, selected in XSPEC with
\texttt{abund wilm}. We ignore channels flagged as bad and group the spectra
for use with the $\chi^2$ statistic. The Galactic
column is fixed at $N_{\rm H}=9.97\times10^{20}\,{\rm cm^{-2}}$ and the
redshift is fixed at $z=0.018$. We do not allow an additional intrinsic
absorber to vary between epochs, since this would introduce an unconstrained
source of apparent flux variability.

The table spectra are calculated for $D_0=300\,{\rm Mpc}$. We therefore fix
the normalization at
$N_{\rm tab}=(D_0/D_{\rm L})^2\simeq(300/78)^2=14.7$. The black-hole mass is
linked across the 16 Swift spectra. The accretion
rate and $\beta$ remain free for each epoch. This enforces a common physical
black-hole mass while testing whether the changing continuum can be described
by the disc-corona parameters alone.

\subsection{Spectral fitting results}
\label{sec:fitting_results}

\begin{figure*}
  \centering
  \includegraphics[height=0.92\textheight,width=\textwidth,keepaspectratio]{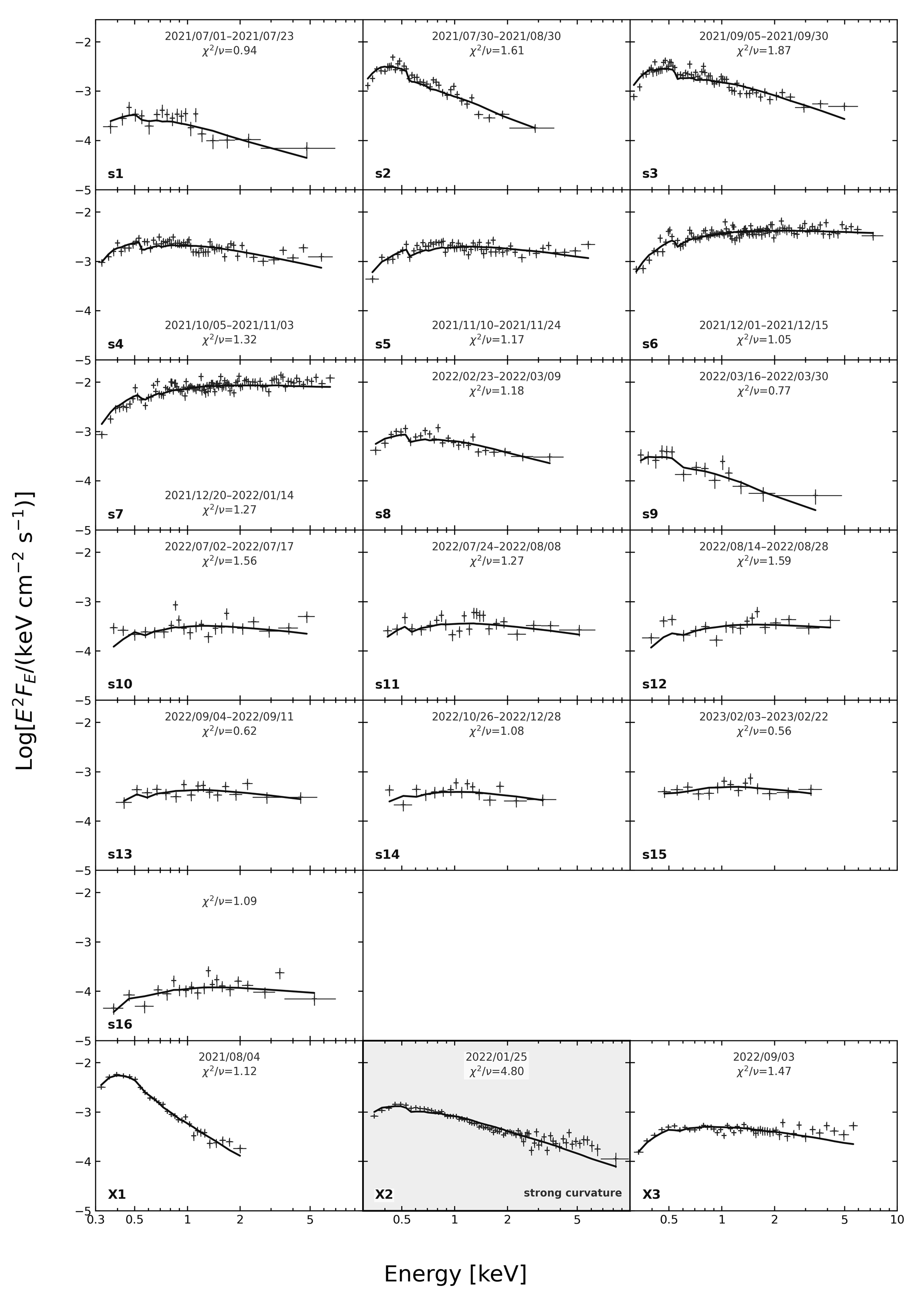}
  \caption{Unfolded Swift/XRT spectra s1--s16 and XMM-Newton spectra X1--X3
  of AT2021ehb with the best-fitting slim disc-corona models. Each panel shows
  $\log[E^2F_E/({\rm keV\,cm^{-2}\,s^{-1}})]$. The dates and reduced
  $\chi^2$ values are shown in each panel. The three XMM-Newton spectra are
  fitted at the common fixed mass $\log(M_{\rm BH}/M_\odot)=6.317$. X1 and X3
  are broadly described, whereas the coherent data--model mismatch in the
  shaded X2 panel exposes additional spectral complexity that the current
  smooth disc-corona model cannot describe.}
  \label{fig:at2021ehb_spectra}
\end{figure*}

The unfolded spectra and the best-fitting models are shown in
Fig.~\ref{fig:at2021ehb_spectra}, and the fitted parameters are listed in
Table~\ref{tab:slimdc_fit}. The model captures the broad Swift/XRT evolution.
The data move between relatively soft spectra and spectra in which a strong
coronal tail masks much of the disc emission. This behaviour requires changes
in both $\mdotd$ and $\beta$; it cannot be reproduced by changing only the
normalization of a fixed spectral shape.

The joint Swift fit gives
$\log(M_{\rm BH}/M_\odot)=6.32\pm0.06$, with
$\chi^2/{\rm d.o.f.}=787.00/593$. The model describes the overall sequence,
although the low null-hypothesis probability and the residuals in several
bright stacks show that it is not a complete description of every epoch. In
particular, the very small fitted $\beta$ values in s6 and s7 indicate that
these spectra require very strong coronal heating within the present model.

Fitting the three XMM-Newton spectra with the mass fixed at the Swift value
gives a joint statistic of
$\chi^2/{\rm d.o.f.}=476.24/148$. X1 and X3 give 31.44/28 and 70.76/48,
respectively, so their broad continua are reasonably described at the common
mass. X2 instead gives 374.05/78. Its residuals show coherent curvature
across the band rather than random statistical scatter. We retain X2 because
this failure is physically informative: it shows additional spectral complexity
that cannot be absorbed into a change of $\mdotd$ or $\beta$. It does not imply
that the black-hole mass changed between epochs. Broad reflection, an ionized
outflow or a more complex corona can modify the continuum
\citep{Yao2022AT2021ehb,Xiang2024AT2021ehb}, but these components are not
included in the present model fits.

For comparison, Table~\ref{tab:slimdc_fit} also lists a fit to the three
XMM-Newton spectra in which the linked mass was free. This fit gives
$\log(M_{\rm BH}/M_\odot)=6.92\pm0.05$ and
$\chi^2/{\rm d.o.f.}=408.9/147$, with individual statistics of 112.7/28,
228.2/78 and 68.0/48 for X1--X3. Allowing the mass to vary therefore
redistributes the mismatch between X1 and X2 rather than removing the
spectral complexity. We retain this solution as a diagnostic comparison, not
as an independent precision mass measurement.

X1 is the softest spectrum and may therefore be the most disc-dominated.
Fitting X1 separately gives $\log(M_{\rm BH}/M_\odot)=6.3\pm0.1$, although
this estimate is model-dependent because the spectrum is still clearly
Comptonised.

\begin{table*}
  \centering
  \caption{Best-fitting slim disc-corona parameters for AT2021ehb. The
  black-hole mass is linked across the Swift/XRT spectra. For XMM-Newton we
  list both the original joint fit with a linked free mass and the fits with
  the mass fixed at the Swift value. The accretion rate and magnetic parameter
  are free for each spectrum. The quoted uncertainties are the formal XSPEC
  errors.}
  \label{tab:slimdc_fit}
  \small
  \renewcommand{\arraystretch}{1.05}
  \begin{tabular*}{0.96\textwidth}{@{\extracolsep{\fill}}llccc@{}}
    \toprule
    ID & Date range & $\log\mdotd$ & $\beta$ & $\chi^2/{\rm dof}$ \\
    \midrule
    \multicolumn{5}{c}{Swift/XRT: $\log(M_{\rm BH}/M_\odot)=6.317\pm0.055$} \\
    \midrule
    s1  & 2021-07-01--2021-07-23 & $-1.138\pm0.055$ & $176.28\pm12.11$ & 18.71/20 \\
    s2  & 2021-07-30--2021-08-30 & $-0.414\pm0.016$ & $201.00\pm13.41$ & 61.15/38 \\
    s3  & 2021-09-05--2021-09-30 & $-0.482\pm0.018$ & $98.63\pm3.63$ & 119.42/64 \\
    s4  & 2021-10-05--2021-11-03 & $-0.637\pm0.035$ & $51.74\pm2.87$ & 79.20/60 \\
    s5  & 2021-11-10--2021-11-24 & $-0.807\pm0.053$ & $30.88\pm3.15$ & 67.97/58 \\
    s6  & 2021-12-01--2021-12-15 & $-0.580\pm0.166$ & $0.100\pm0.447$ & 110.00/105 \\
    s7  & 2021-12-20--2022-01-14 & $-0.246\pm0.132$ & $0.100\pm0.378$ & 134.57/106 \\
    s8  & 2022-02-23--2022-03-09 & $-0.911\pm0.036$ & $105.48\pm9.89$ & 30.60/26 \\
    s9  & 2022-03-16--2022-03-30 & $-1.045\pm0.069$ & $294.25\pm25.60$ & 11.60/15 \\
    s10 & 2022-07-02--2022-07-17 & $-1.545\pm0.093$ & $43.63\pm14.48$ & 35.93/23 \\
    s11 & 2022-07-24--2022-08-08 & $-1.457\pm0.057$ & $52.98\pm10.62$ & 30.40/24 \\
    s12 & 2022-08-14--2022-08-28 & $-1.610\pm0.080$ & $19.81\pm21.18$ & 28.60/18 \\
    s13 & 2022-09-04--2022-09-11 & $-1.396\pm0.073$ & $49.95\pm11.53$ & 10.47/17 \\
    s14 & 2022-10-26--2022-12-28 & $-1.364\pm0.104$ & $63.61\pm15.43$ & 17.22/16 \\
    s15 & 2023-02-03--2023-02-22 & $-1.323\pm0.094$ & $50.00\pm13.36$ & 8.35/15 \\
    s16 & 2023-12-22--2024-02-19 & $-1.990\pm0.070$ & $31.39\pm25.99$ & 22.81/21 \\
    \midrule
    \multicolumn{5}{c}{XMM-Newton: $\log(M_{\rm BH}/M_\odot)=6.915\pm0.053$ (free)} \\
    \midrule
    X1 & 2021-08-04 & $-0.230\pm0.009$ & $409.14\pm12.01$ & 112.68/28 \\
    X2 & 2022-01-25 & $-0.902\pm0.012$ & $153.07\pm1.96$ & 228.19/78 \\
    X3 & 2022-09-03 & $-1.642\pm0.036$ & $64.66\pm5.42$ & 68.02/48 \\
    \midrule
    \multicolumn{5}{c}{XMM-Newton: $\log(M_{\rm BH}/M_\odot)=6.317$ (fixed)} \\
    \midrule
    X1 & 2021-08-04 & $-0.225\pm0.006$ & $342.365\pm15.009$ & 31.44/28 \\
    X2 & 2022-01-25 & $-0.747\pm0.004$ & $118.519\pm1.524$ & 374.05/78 \\
    X3 & 2022-09-03 & $-1.313\pm0.030$ & $58.476\pm4.880$ & 70.76/48 \\
    \bottomrule
  \end{tabular*}
\end{table*}

\subsection{Evolution of fitted parameters}
\label{sec:fitted_parameter_evolution}

\begin{figure*}
  \centering
  \includegraphics[width=0.70\textwidth]{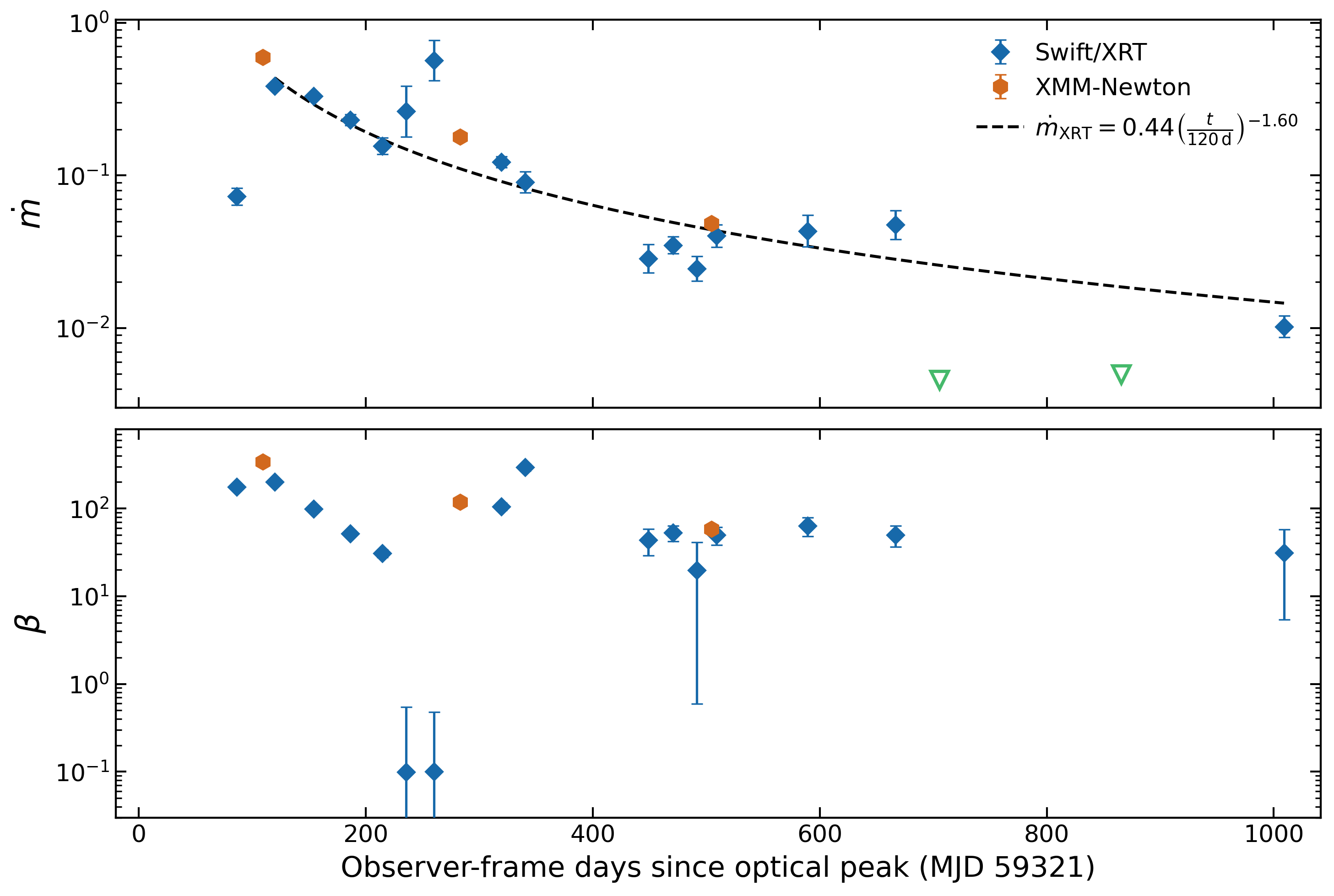}
  \caption{Evolution of the X-ray inner-flow accretion-rate proxy (top) and magnetic parameter
  (bottom) for AT2021ehb. Blue diamonds show Swift/XRT measurements and orange
  hexagons show XMM-Newton measurements; the XMM points use the common fixed
  black-hole mass and their
  formal parameter uncertainties are not shown. Green open downward triangles
  in the upper panel mark the count-rate-derived 90 per cent upper limits on
  $\mdotd_{\rm XRT}$ inferred from the combined Swift observations
  00014217065--00014217066 and from observation 00014217067. These upper
  limits are not independent spectral fits and are excluded from the dashed
  power-law fit. Their low photon counts do not independently constrain
  $\beta$, so no corresponding points are shown in the lower panel. The dashed
  line describes the main Swift accretion-rate
  decline. Smaller $\beta$
  corresponds to a stronger magnetic field relative to the
  gas-plus-radiation pressure.}
  \label{fig:at2021ehb_params}
\end{figure*}

The time evolution of the fitted parameters is shown in
Fig.~\ref{fig:at2021ehb_params}. The Swift values show an overall decline in the
effective inner-flow accretion rate, interrupted by the bright hard interval. A
power-law fit to the main declining sequence gives approximately
\begin{equation}
  \mdotd_{\rm XRT} = 0.44
  \left(\frac{t}{120\,{\rm d}}\right)^{-1.60}.
  \label{eq:mdot_decline}
\end{equation}
This relation describes the parameter controlling the X-ray-emitting inner
flow. 

However, much faster variability in $\dot{m}$ is superimposed on the slow
decline. For example, $\dot{m}$ increases by a factor of 3 from s5--s7
($t=215$--$260$~d), followed by a similarly rapid return. There is an even
larger drop, by a factor of more than 10, during the late-time decline, from
s15 at $t\sim670$~d to the very weak detections
at $t=705$ and $870$~d. 
These points were not included in the power-law fit in
equation~(\ref{eq:mdot_decline}), but they clearly show a deep fading state
before the source recovers to the s16 detection at $t\sim1000$~d, which has
$\mdotd_{\rm XRT}\simeq0.01$.

The non-monotonic $\beta$ evolution shows that the spectral changes are not set
by the accretion rate alone. Both $\beta$ and the effective $\mdotd$ change
sharply during the rapid hard-to-soft transition; their physical implications are discussed
in Section~\ref{sec:rapid_changes}.

\section{Discussion}
\label{sec:discussion}

\subsection{Overall disc-corona evolution}

The AT2021ehb fits show that the disc-corona model developed here captures
most of the source behaviour as it brightens, hardens and fades. However, the
high signal-to-noise XMM-Newton spectra are not always well fitted, showing
that the data contain more complexity than the model includes.

The fitted mass accretion rate nevertheless follows an approximate power-law
decline. \citet{Mummery2026} models the evolution of TDEs using standard
accretion-disc equations without X-ray coronae and predicts the timescale for
the mass accretion rate to decline from $\dot m=1$ to $0.01$. They assume a
$10^6\,M_\odot$ black hole, comparable to the mass estimated here, and find a
timescale of $\sim2000$~d, similar to the $\sim1000$~d observed in this source.

\subsection{Rapid changes in luminosity and hardness}
\label{sec:rapid_changes}

However, the fits also show variability that cannot be captured by this smooth
trend. First, the inferred mass accretion rate increases by a factor of 3 from
$215$ to $260$~d during the rise to the X-ray peak, followed by a similarly
rapid decline. There is an even deeper drop at late times, with $\dot m$ a
factor of 10 lower than predicted by the power-law decline between $700$ and
$870$~d. This rapid variability is difficult to reproduce with viscous disc
evolution alone.

The model also requires rapid changes in the magnetic coupling between the disc
and corona to follow the spectral evolution from disc-dominated at the start
to corona-dominated at the X-ray peak. The largest change in this ratio occurs
between $t=215$ and $235$~d, when $\beta$ decreases by a factor of 300 while
the mass accretion rate changes by less than a factor of 2.

The magnetic field can change on much shorter timescales than the viscous
timescale, so this is not necessarily problematic. It does, however, show that
an additional parameter besides $\dot m$ controls the strength of the corona.

\subsection{Black-hole mass}

The linked Swift/XRT fit gives $\log(M_{\rm BH}/M_\odot)=6.317\pm0.055$, and we
use this same mass when testing all three XMM-Newton spectra. The host stellar
velocity dispersion gives
$\log(M_{\rm BH}/M_\odot)=7.03\pm0.15\ ({\rm stat})\pm0.29\ ({\rm sys})$ for the
adopted $M_{\rm BH}$--$\sigma$ relation \citep{Yao2022AT2021ehb}. The X-ray and
host estimates therefore place the source broadly in the
$10^6$--$10^7\,M_\odot$ range once model dependence and the scatter of the
scaling relation are considered.

The free-mass XMM fit in Table~\ref{tab:slimdc_fit} illustrates why a more
precise claim is not justified. X2 alone can yield an apparent continuum mass
near $10^5\,M_\odot$ as the model tries to reproduce its curvature.
\citet{Xiang2024AT2021ehb} likewise inferred
a characteristic mass of $10^5$--$10^6\,M_\odot$ from six XMM-Newton
observations obtained immediately after our X2 epoch, during the same
intermediate, post-drop interval. They did not fit the X2 spectrum used here. The
agreement between these low apparent masses is therefore a comparison between
neighbouring observations, not a repeat fit to the same data.

The true black-hole mass cannot change between epochs. These shifts instead
show that continuum-based mass estimates are unstable when strong
Comptonization and additional spectral components obscure the disc. The result
depends on the assumed disc-corona geometry, colour correction, reflection and
the conversion between fitted emitting area and inner-disc scale. We therefore
regard $M_{\rm BH}\sim10^6$--$10^7\,M_\odot$ as a reasonable scale for this
application, not as a precision continuum-fitting measurement.

\subsection{Comparison with AGNs and X-ray binaries}

The model explored here is a magnetically heated corona above a standard
vertically averaged Shakura--Sunyaev disc, modified to include optically thick
advection above the Eddington limit. The corona is powered by magnetic heating,
parameterised by the ratio of disc pressure to magnetic pressure, $\beta$, which
is taken to be constant with radius. This prescription determines the
self-consistent optical depth and temperature of the corona at each radius.
The radial variation is relatively slow and smooth, while the gravitational
power release increases strongly towards smaller radii. Consequently, while there is radial structure, most of
the spectrum is determined by conditions within the inner $10R_{\rm S}$, and
the integrated spectra resemble a standard disc with a single thermal-Compton component with the
optical depth and temperature evaluated over $5$--$10R_{\rm S}$.

These solutions differ from observed AGN spectra over comparable
black-hole mass and accretion-rate ranges. AGN spectra generically show an upturn below $2~\mathrm{keV}$ relative to an extrapolation of the hard coronal power law, commonly referred to as the ``soft X-ray excess''. It is often modelled
as a warm, optically thick Comptonising region additional to the hot, optically thin corona giving an emission component which spans between the UV and soft X-ray regions \citep{Petrucci2018,Kubota2018,Jin2012,Mehdipour2011}. 
This component is particularly prominent at high
accretion rates ($\dot{m}\sim 1$, as at the peak here) in narrow-line Seyfert 1
galaxies (NLS1s), which typically host black holes with
$\log(M/M_\odot)\sim 6$--$7$, but it is not present in our model. To form this component requires more heating towards the top of the disc than expected in the simple Shakura--Sunyaev disc viscous heating prescription used here (see Section 2.2, e.g. \citealt{Petrucci2020}).

Another clear difference between this TDE and AGN is that the TDE has spectra around the peak at $\dot{m}\sim 1$ which are dominated by the hot corona (e.g. s6--s7 in Fig.~\ref{fig:at2021ehb_spectra}).
AGN spectra at similar $\dot{m}$ generally have weak coronal emission, parameterised by a large
bolometric correction $\kappa_{2\text{--}10\,{\rm keV}}=L_{\rm bol}/L_{2\text{--}10\,{\rm keV}}=30\text{--}100$ \citep{Jana26}. By contrast, our TDE fits show s7 has $\kappa_{2\text{--}10\,{\rm keV}}=7$, showing the hot corona is dominant rather than weak as in AGN at similarly high $\dot{m}$.

Other spectra around the peak of this TDE have only slightly smaller $\dot{m}$ yet they show much weaker coronae, e.g. s2 has $\dot{m}\sim 0.4$ compared to $0.6$ for s7 but the model has a bolometric correction of $\kappa_{2\text{--}10\,{\rm keV}}\sim 500$. These are more analogous to those seen in similarly bright AGN, but the rapid change in coronal power from s2--s3 versus s4--s7 is not seen in AGN. Our model can follow this evolution by a rapid change in magnetic coupling, and this can happen on much faster timescales than the viscous timescale of the thin disc. 

XRB do show rapid changes in coronal power at similarly high $\dot{m}$, especially around the peak of the rapid rise to
outburst in transient XRBs. XRB around their peak have complex spectra and variability,
plausibly because the accretion rate changes too rapidly for the flow to
remain in steady state and/or the magnetic structures are changing. TDEs may similarly be out of steady state, making XRBs a
better analogy than AGNs. 

However, both XRB and AGN show a 
transition at low luminosities
to a corona-dominated hard state, typically with
$\Gamma\sim 1.5$--$1.8$ \citep{Remillard2006,Done2007,Hagen2024}.  Increasing coronal
dominance has also been interpreted as evidence for a state transition in
TDEs, notably AT2018fyk \citep{Wevers2021Rapid3}. However, the power-law
indices in `hard' state TDEs
remain close to
$\Gamma\sim 2$, with some dependence on the assumed soft spectral component
\citep[their table~4]{Wevers2021Rapid3}.
The dimmest spectra for our data here (AT2021ehb) are likewise clearly corona-dominated, but
with $\Gamma\sim 2$. They therefore do not show the harder spectra expected
at the transition, even though they appear to have $\dot{m}\sim 0.01$.
Avoiding this soft-to-hard transition would require the black-hole mass to be
overestimated, so that $\dot{m}$ was underestimated. However, this would push
the inferred mass further from that expected from the $M$--$\sigma$ relation,
creating another inconsistency.

Nevertheless, experience from XRBs shows that spectral estimates of black-hole
mass (and spin) are strongly model dependent when the spectra are
corona-dominated, but more robust in disc-dominated states (e.g. \citealt{Done2007}). The softest
spectrum in this TDE outburst, X1, gives $\log M\sim 6.3$, so $\dot{m}$ spans
$1$--$0.01$ as before.

Another potentially relevant class for comparison to TDEs is the ultraluminous X-ray sources (ULXs).
These are highly super-Eddington flows from a high-mass companion star onto either a
neutron star, as in pulsating ULXs, or a stellar-mass black hole. ULXs can show
complex spectra and variability because of obscuration and reflection from a
super-Eddington wind or a puffed-up inner-disc funnel, in addition to their
intrinsic spectral evolution with $\dot m$ (e.g. \citealt{Middleton2026}). Disc winds and variable obscuration
have likewise been invoked to explain complex X-ray spectra and variability in
extreme NLS1s \citep{hagino2016,parker2018,jin2023}.

Alternatively, TDEs may be inherently different from other known
accretion-powered black-hole systems. They are transient, non-steady and
possibly misaligned flows in which radiation trapping, fallback, disc tearing,
winds and reprocessing can alter the spectral evolution.

\section{Conclusions}
\label{sec:conclusions}

We have constructed a colour-corrected, magnetically heated slim disc-corona
model for TDE X-ray spectra and applied it to AT2021ehb. Our main conclusions
are:
\begin{enumerate}
\item We solve a unified slim disc-corona model in which gas pressure,
radiation pressure, magnetic pressure, advection and coronal cooling are treated
together. This gives continuous radial structures across pressure regimes and
allows the corona to vary smoothly with accretion
rate for a fixed magnetic coupling parameter. This captures the broad X-ray spectral evolution of AT2021ehb. 
\item The fits favour a black-hole mass scale of
$M_{\rm BH}\sim10^6$--$10^7\,M_\odot$. This is physically reasonable but not a
precision measurement: strong Comptonization and additional continuum
structure make the inferred mass sensitive to the assumed disc-corona geometry.
\item The overall mass accretion rate declines from $\dot m=1$ to $0.01$ over
1000~d, roughly consistent with the expected viscous-disc timescale evolution.
\item More rapid changes in $\dot m$ are superimposed on this trend and require
additional processes such as inhomogeneous fallback.
\item Rapid changes in magnetic heating are also required to follow the large
change in coronal dominance around the peak.
\item We compare the behaviour with AGNs and XRBs and show that transient XRB
outbursts can produce similarly large changes in coronal dominance around their
peak. However, both AGNs and XRBs show a clear transition below
$\dot m\sim0.01$ to a much harder corona ($\Gamma\sim1.8$) than is seen here
and in other TDEs. This suggests that some aspects of TDEs are specific to
these transient, non-steady flows.

\end{enumerate}

\section*{Acknowledgements}
WC thanks Chenlei Guo for helpful discussions and sharing
experience on the construction of the disc-corona model. 
EQ acknowledges funding from the
National Key R\&D Program of China
(grant no. 2023YFA1607903), the Strategic Priority Research Programme of the Chinese Academy
of Sciences (grant no. XDB0550200) and the National Natural Science Foundation
of China (grant no. 12333004). CD acknowledges support from STFC through the
grant ST/T000244/1, and WC acknowledges support from the China Scholarship
Council (CSC).

\section*{Data Availability}
The Swift/XRT and XMM-Newton data used in this work are publicly available
from their respective mission archives. The XSPEC additive table model
\texttt{slimdc.fits}, the sorted source spectral grid \texttt{slimdc.csv}, and
usage documentation are publicly available from Zenodo at
\href{https://doi.org/10.5281/zenodo.22015074}{doi:10.5281/zenodo.22015074}.

\appendix
\section{Additional model calculations}
\label{app:model_extra}

This appendix shows the same model diagnostics as in
Section~\ref{sec:model_results}, but for $M_{\rm BH}=10^5\,\msun$ and
$10^7\,\msun$.

\begin{figure*}
\centering
\includegraphics[width=0.48\textwidth,height=0.27\textheight,keepaspectratio]{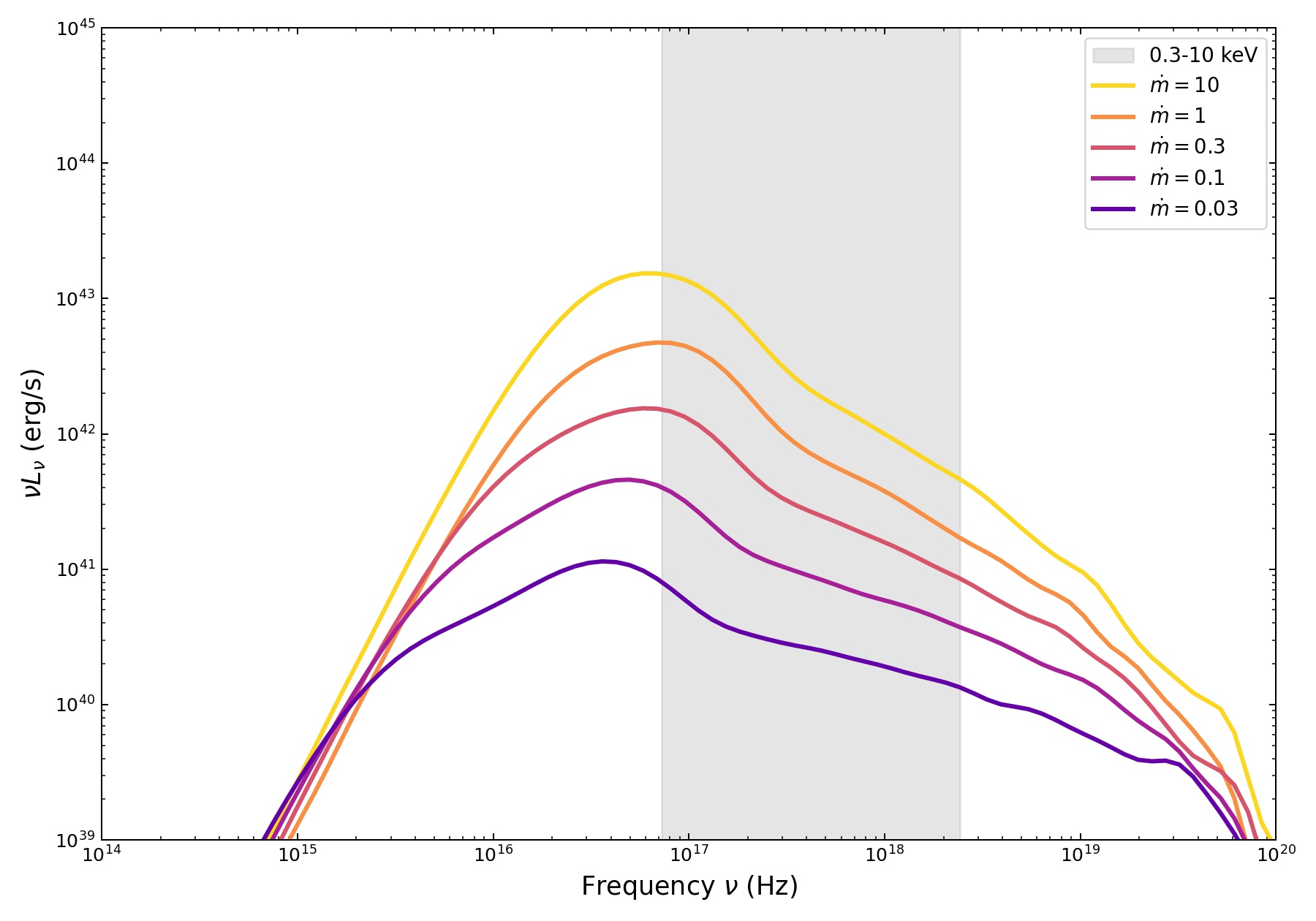}\hfill
\includegraphics[width=0.48\textwidth,height=0.27\textheight,keepaspectratio]{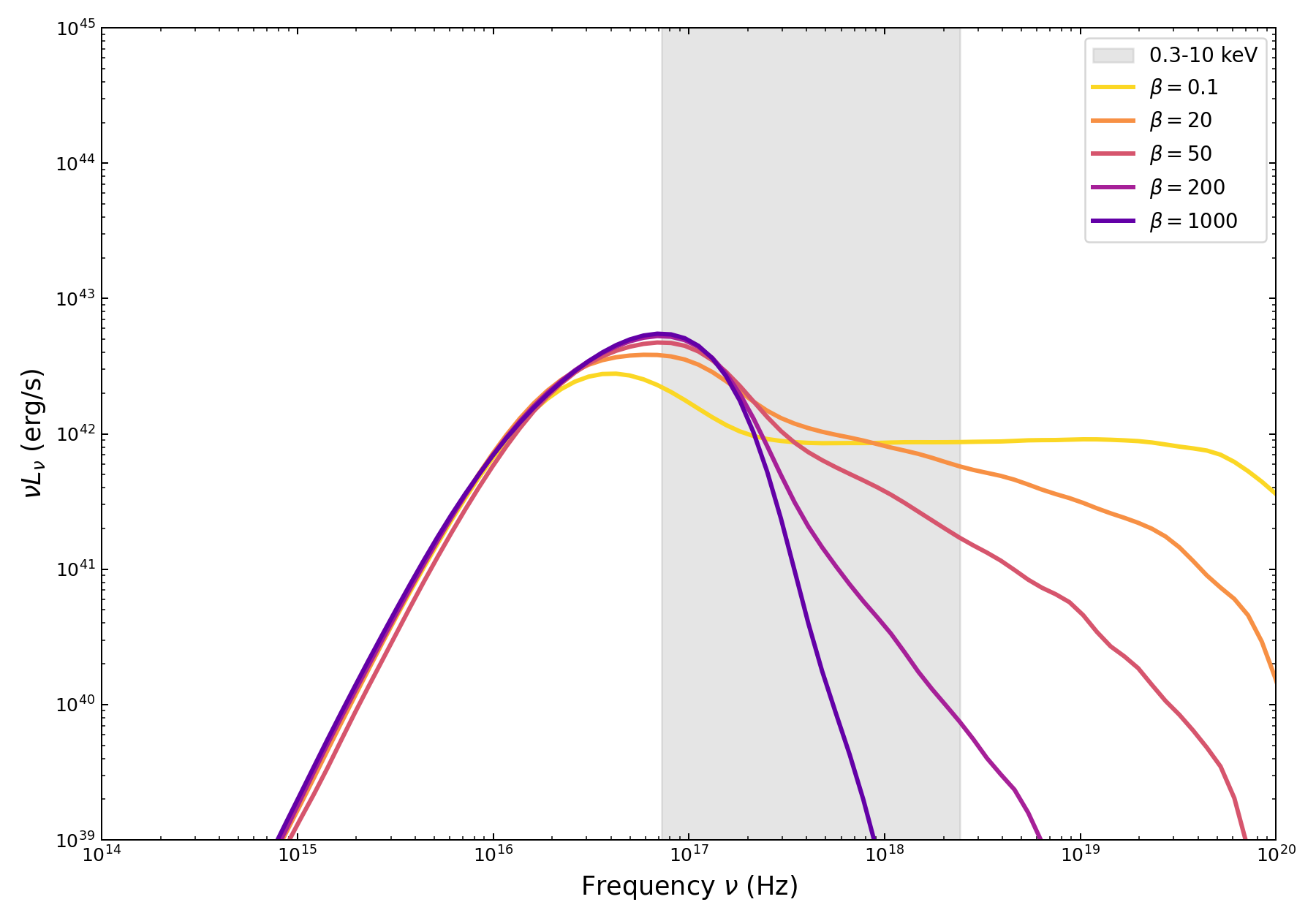}
\caption{Emergent spectra for $M_{\rm BH}=10^5\,\msun$. Left: varying accretion rate at fixed $\beta=50$. Right: varying magnetic parameter at fixed $\mdotd=1$. Smaller $\beta$ corresponds to a stronger magnetic field and a larger coronal energy fraction.}
\label{fig:app_mdot_spectrum_1e5}
\label{fig:app_beta_spectrum_1e5}
\vspace{5mm}
\includegraphics[width=0.86\textwidth,height=0.46\textheight,keepaspectratio]{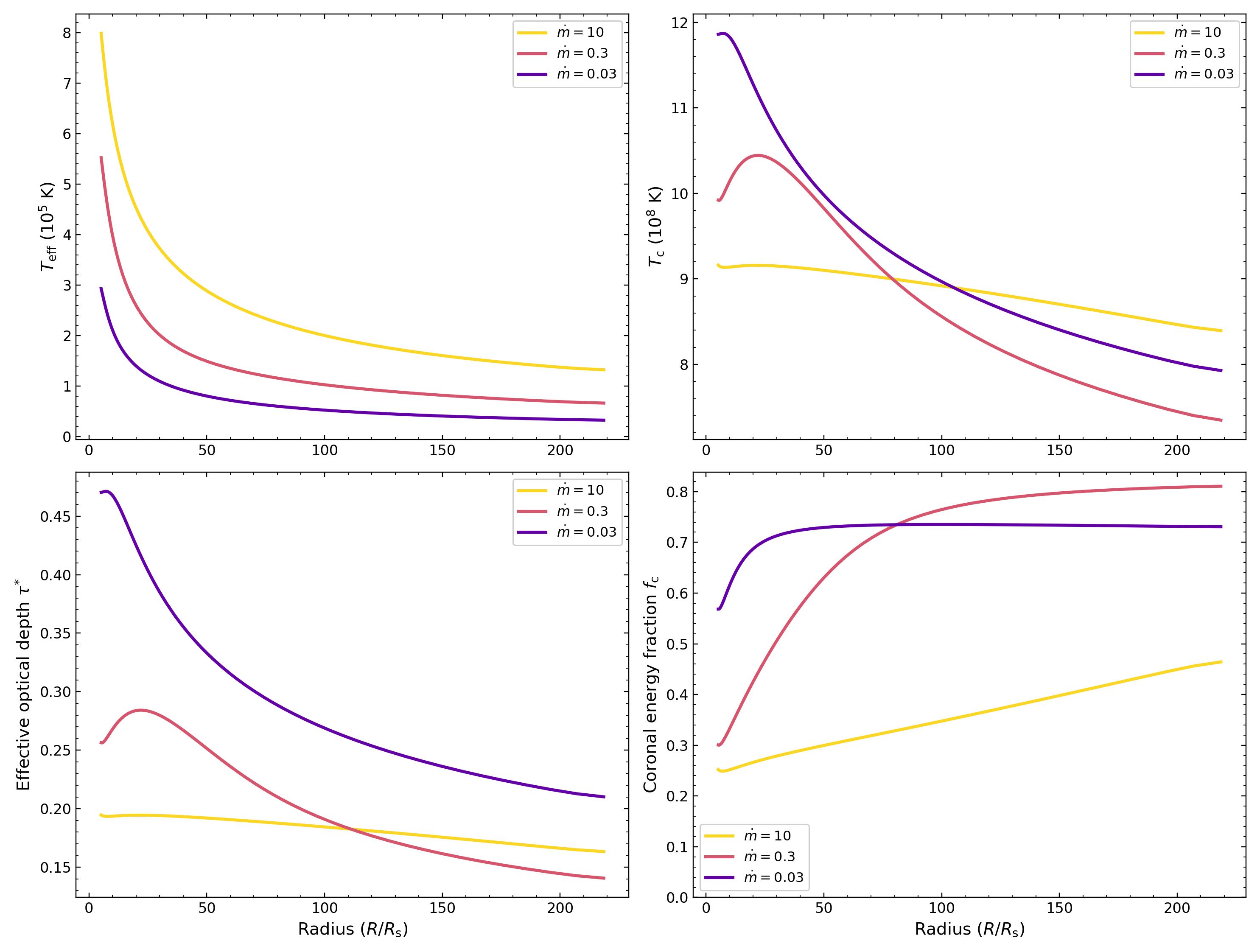}
\caption{Radial structures for $M_{\rm BH}=10^5\,\msun$ and $\beta=50$ at
  representative accretion rates. The panels show the effective temperature
  $\teff$, coronal temperature $\tc$, effective optical depth $\tau^{*}$, and
  coronal energy fraction $f_{\rm c}$.}
  \label{fig:app_mdot_structure_1e5}
\end{figure*}
\clearpage

\begin{figure*}
\centering
\includegraphics[width=0.86\textwidth,height=0.46\textheight,keepaspectratio]{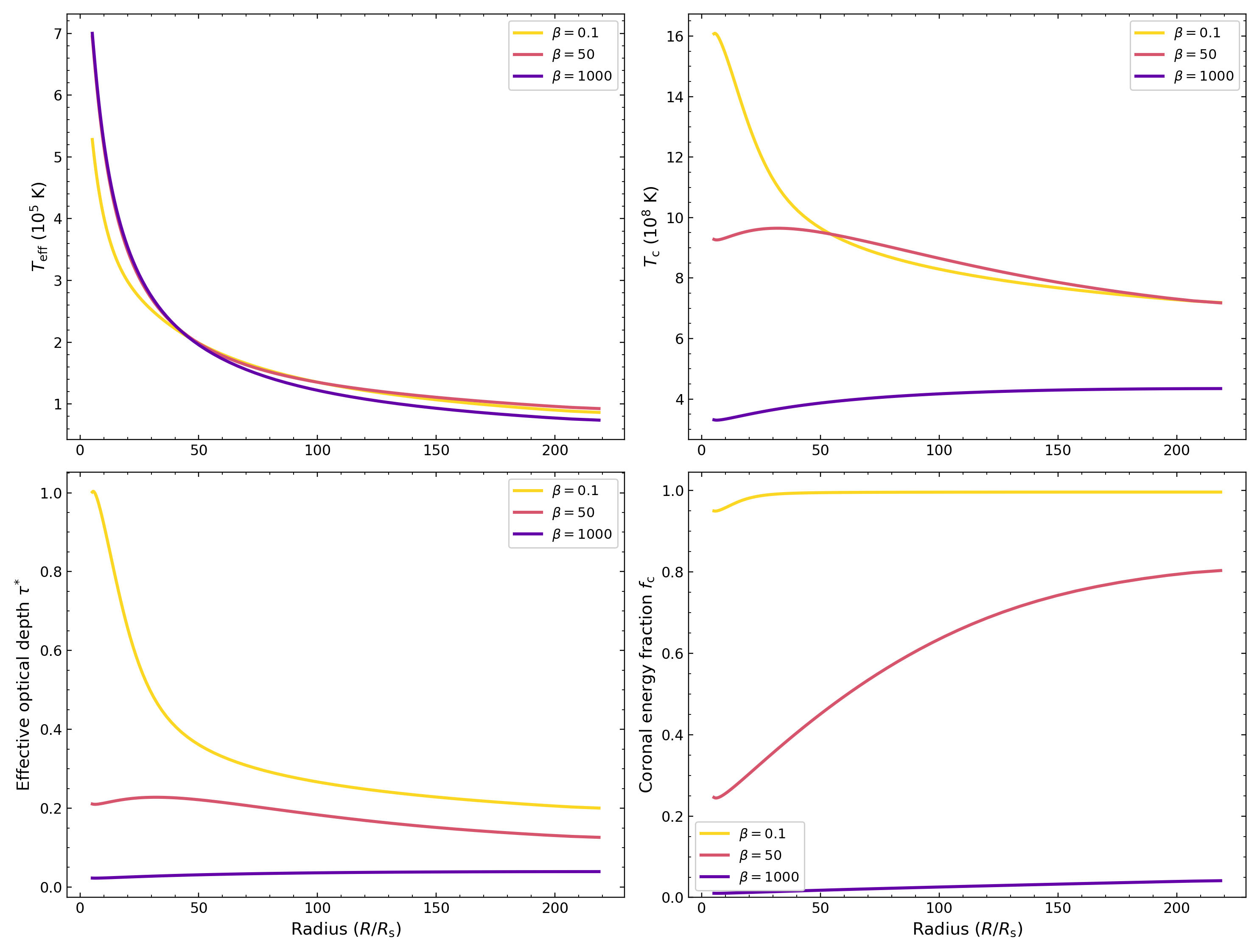}
\caption{Radial structures for $M_{\rm BH}=10^5\,\msun$ and $\mdotd=1$ at
  representative magnetic parameters. The coronal temperature, effective
  optical depth and coronal energy fraction vary strongly with $\beta$, while
  the disc effective temperature changes more weakly.}
  \label{fig:app_beta_structure_1e5}
\vspace{5mm}
\includegraphics[width=0.48\textwidth,height=0.27\textheight,keepaspectratio]{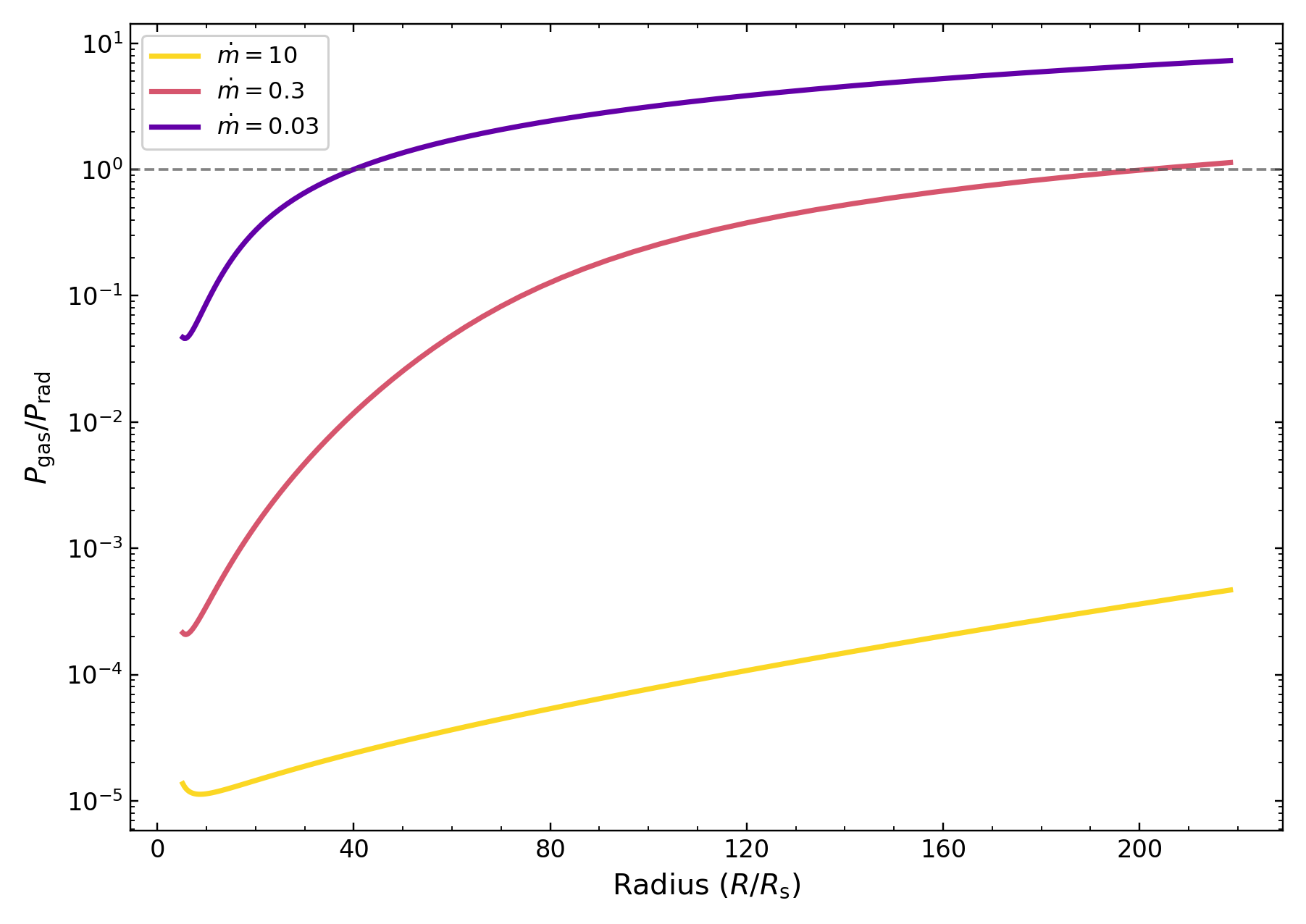}\hfill
\includegraphics[width=0.48\textwidth,height=0.27\textheight,keepaspectratio]{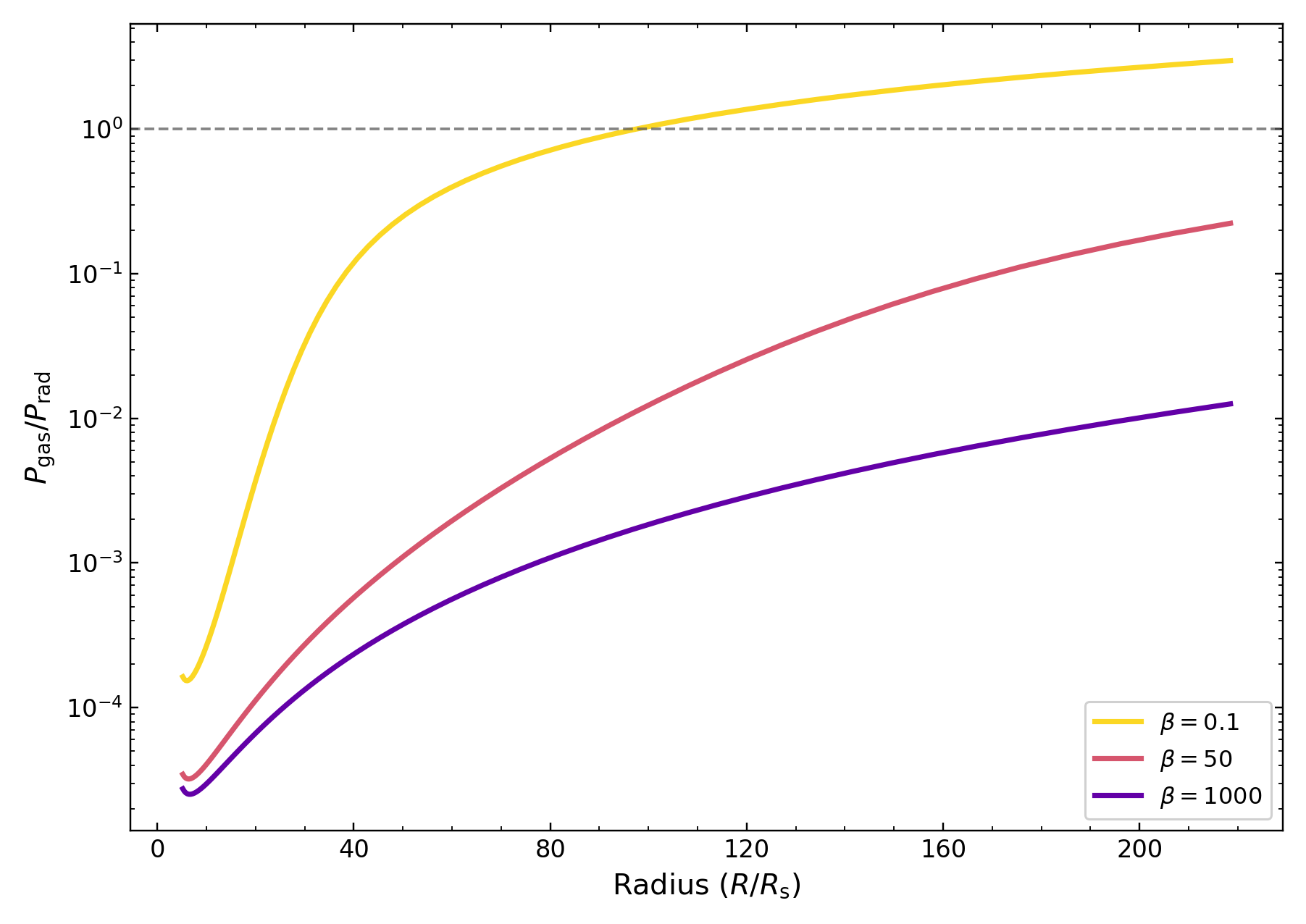}
\caption{Pressure-ratio profiles for $M_{\rm BH}=10^5\,\msun$. Left: representative accretion rates at fixed $\beta=50$. Right: representative magnetic parameters at fixed $\mdotd=1$. The dashed line marks $P_{\rm gas}=P_{\rm rad}$.}
\label{fig:app_mdot_pgas_prad_1e5}
\label{fig:app_beta_pgas_prad_1e5}
\end{figure*}
\clearpage

\begin{figure*}
\centering
\includegraphics[width=0.48\textwidth,height=0.27\textheight,keepaspectratio]{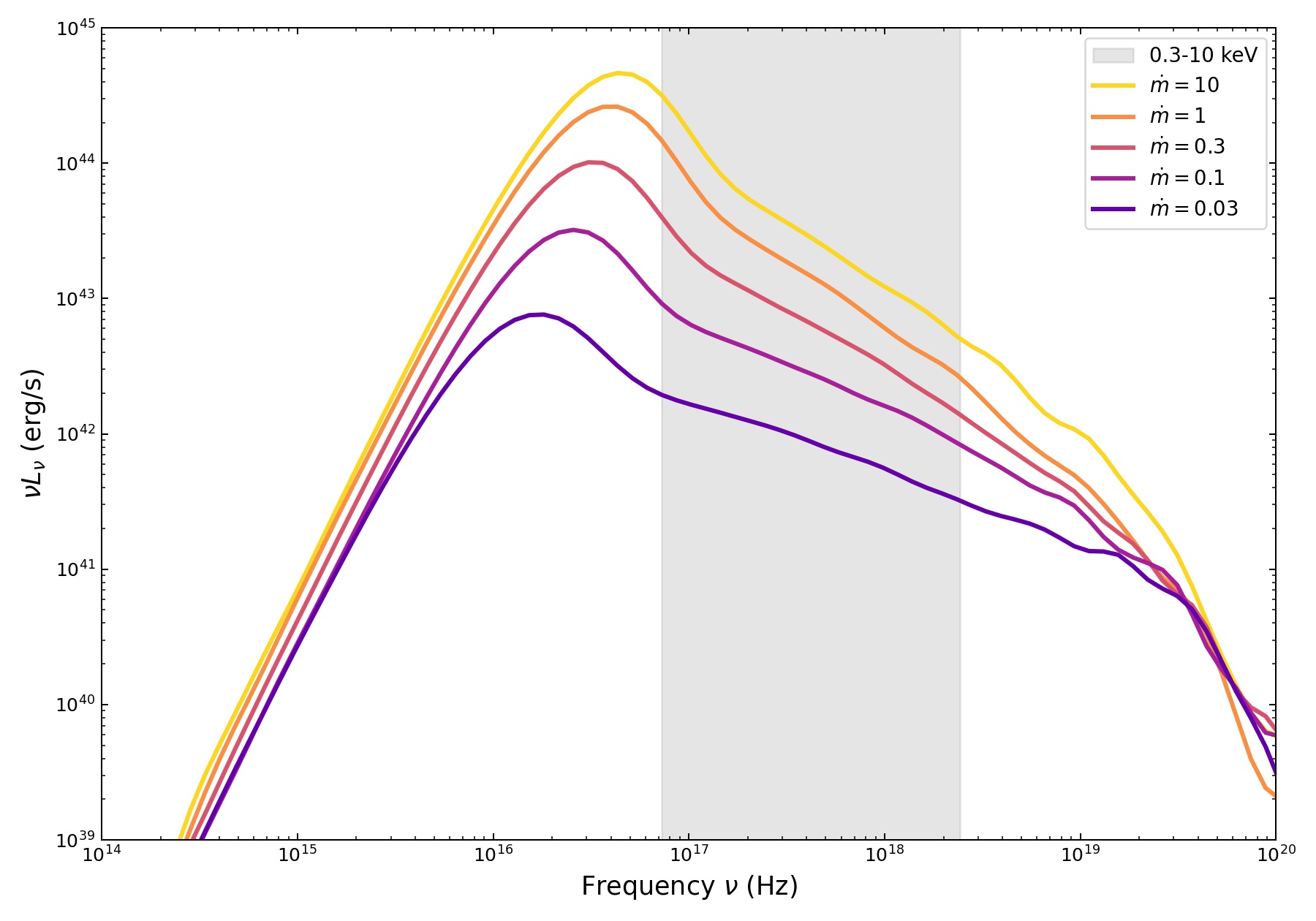}\hfill
\includegraphics[width=0.48\textwidth,height=0.27\textheight,keepaspectratio]{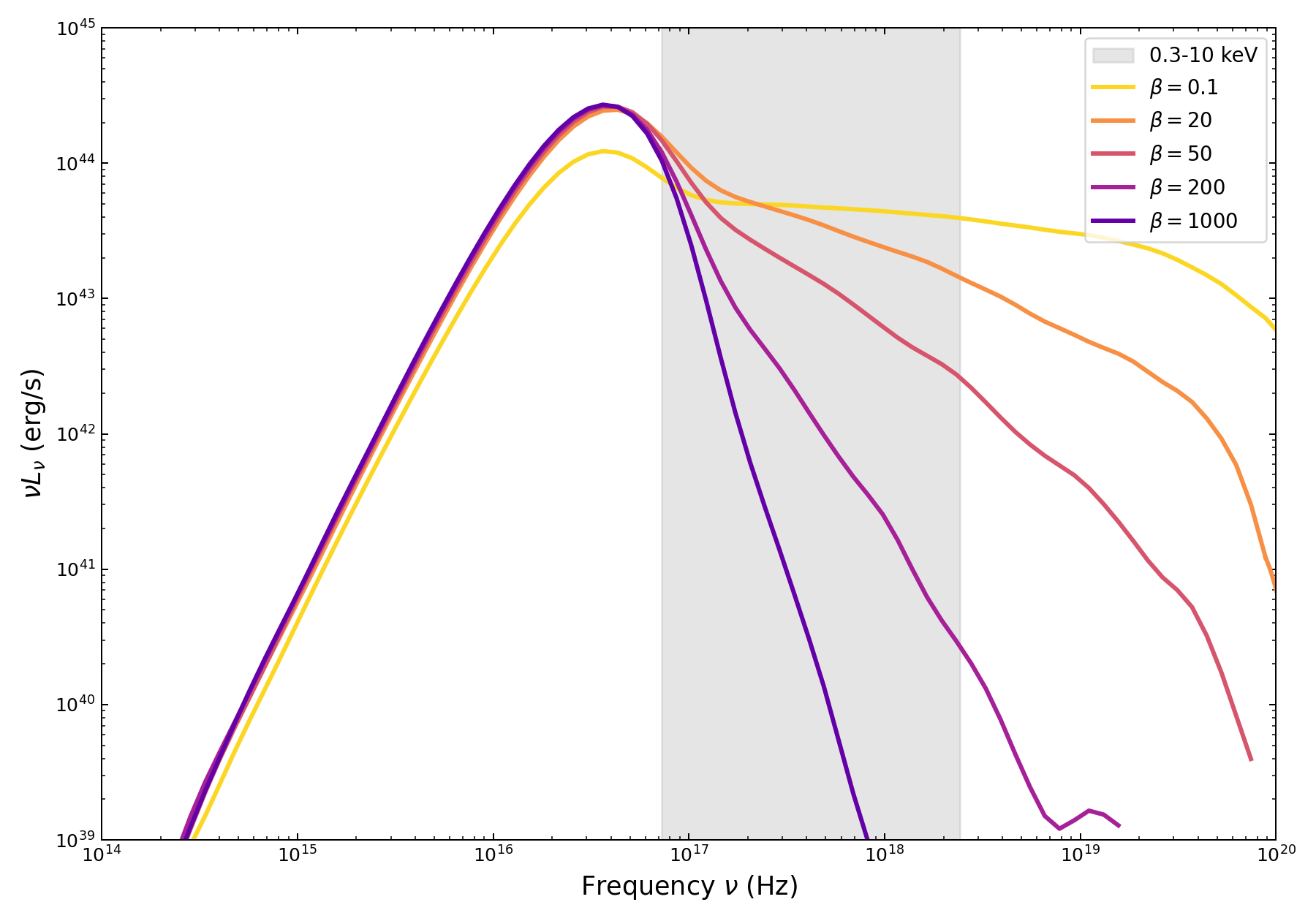}
\caption{Emergent spectra for $M_{\rm BH}=10^7\,\msun$. Left: varying accretion rate at fixed $\beta=50$. Right: varying magnetic parameter at fixed $\mdotd=1$. Smaller $\beta$ corresponds to a stronger magnetic field and a larger coronal energy fraction.}
\label{fig:app_mdot_spectrum_1e7}
\label{fig:app_beta_spectrum_1e7}
\vspace{5mm}
\includegraphics[width=0.86\textwidth,height=0.46\textheight,keepaspectratio]{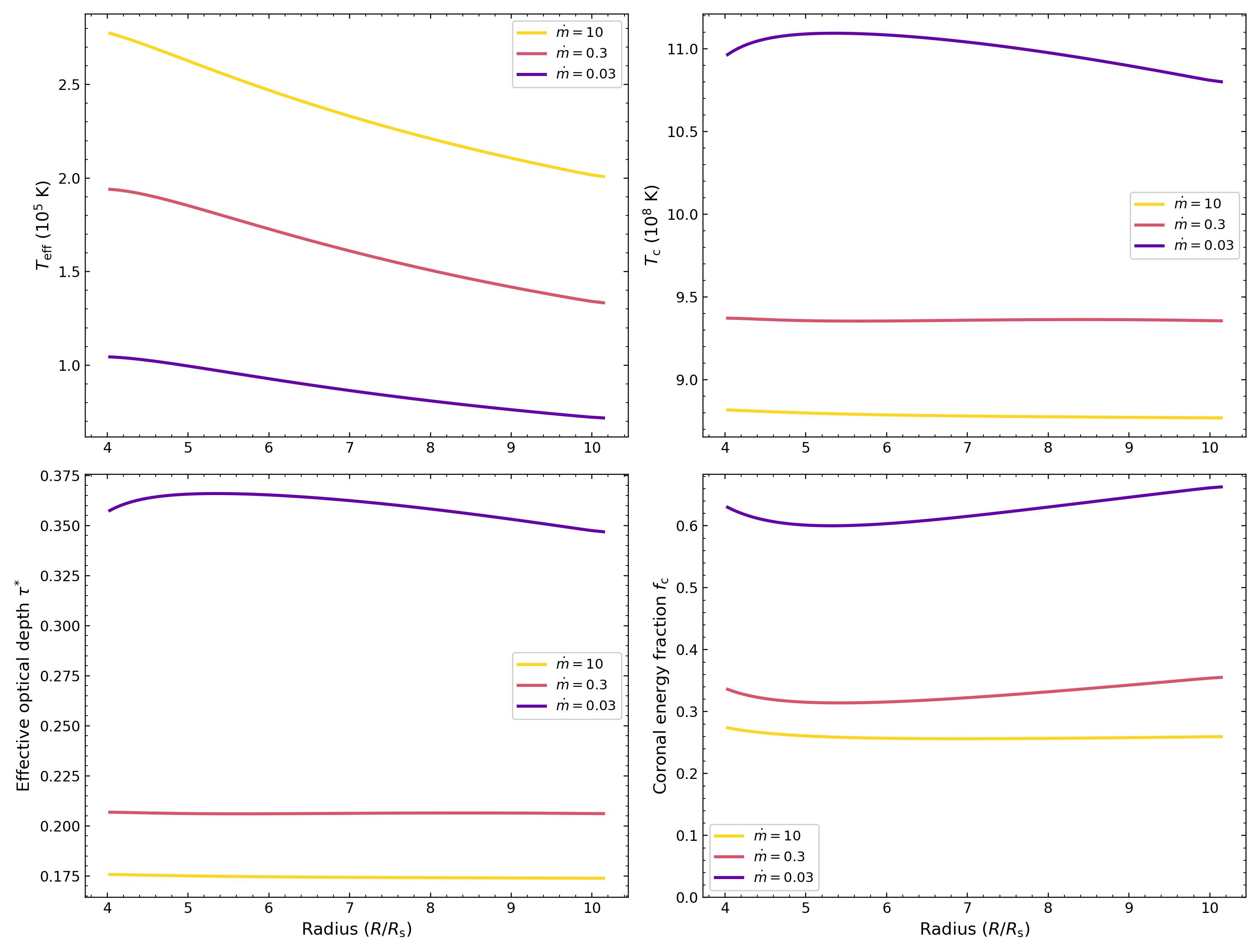}
\caption{Radial structures for $M_{\rm BH}=10^7\,\msun$ and $\beta=50$ at
  representative accretion rates. The panels show $\teff$, $\tc$, the
  effective optical depth $\tau^{*}$, and the coronal energy fraction
  $f_{\rm c}$ for the converged colour-corrected solutions.}
  \label{fig:app_mdot_structure_1e7}
\end{figure*}
\clearpage

\begin{figure*}
\centering
\includegraphics[width=0.86\textwidth,height=0.46\textheight,keepaspectratio]{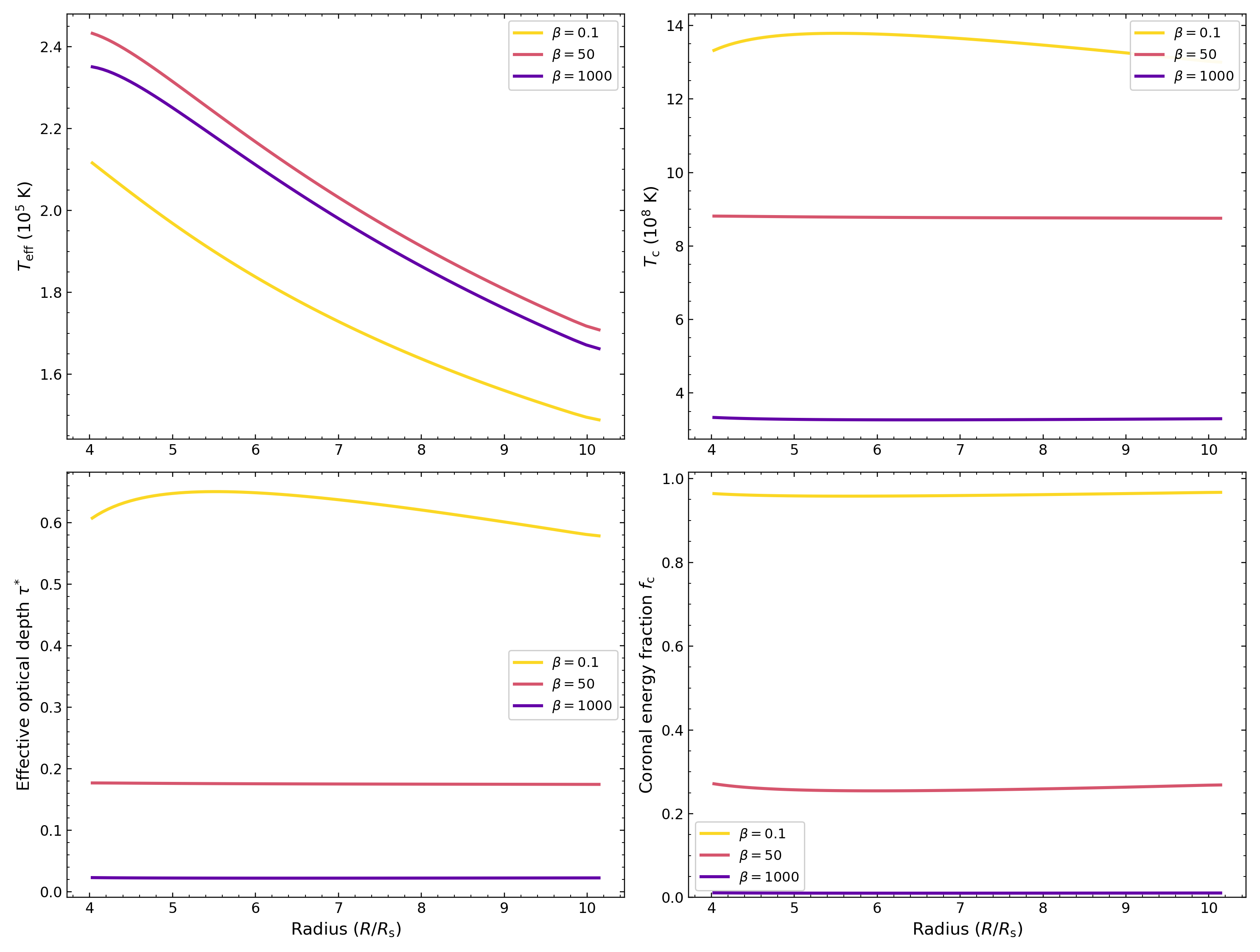}
\caption{Radial structures for $M_{\rm BH}=10^7\,\msun$ and $\mdotd=1$ at
  representative magnetic parameters. The stronger-field solutions show larger
  coronal energy fractions and higher effective optical depths, producing the
  harder spectra shown in the corresponding spectral comparison.}
  \label{fig:app_beta_structure_1e7}
\vspace{5mm}
\includegraphics[width=0.48\textwidth,height=0.27\textheight,keepaspectratio]{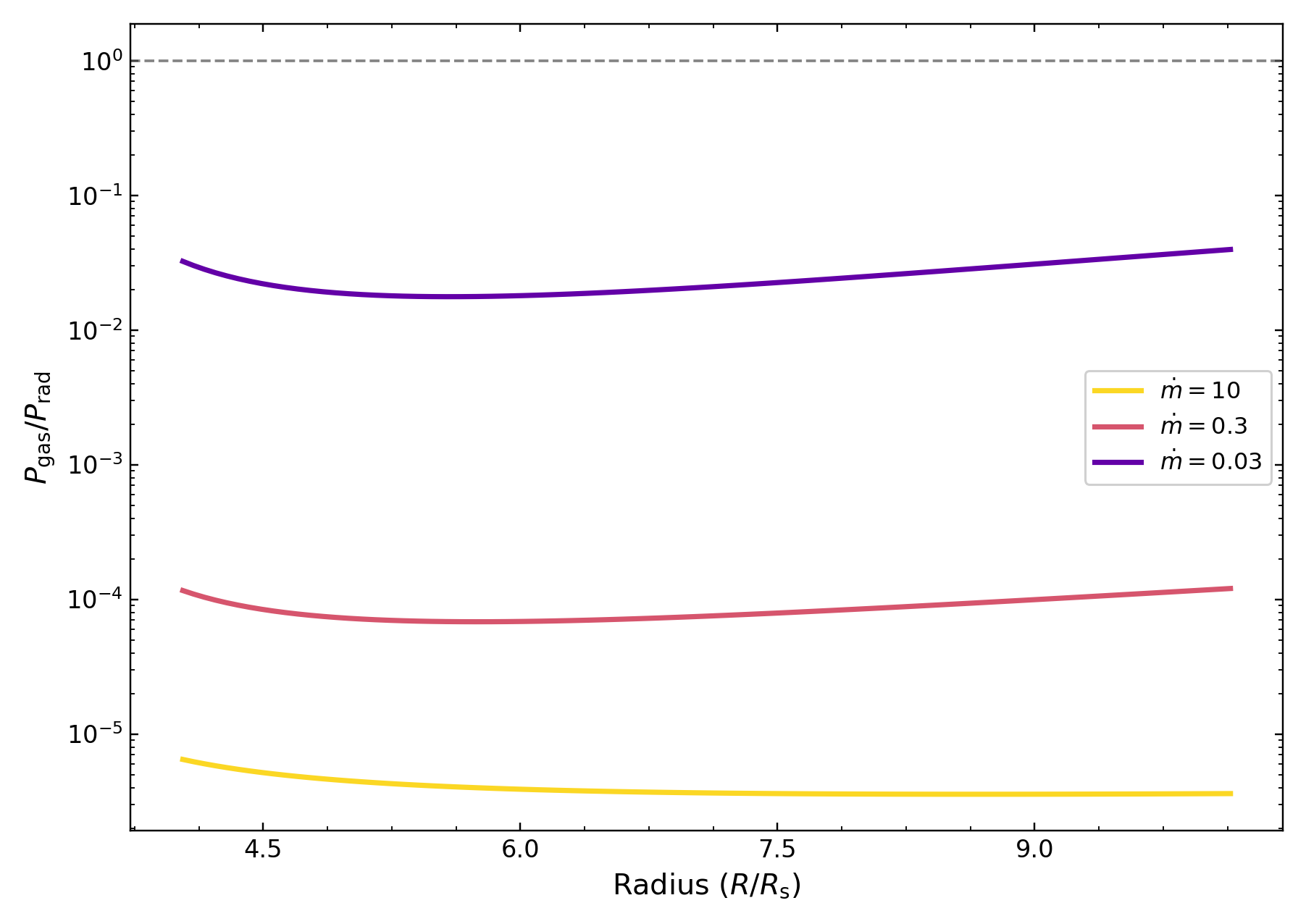}\hfill
\includegraphics[width=0.48\textwidth,height=0.27\textheight,keepaspectratio]{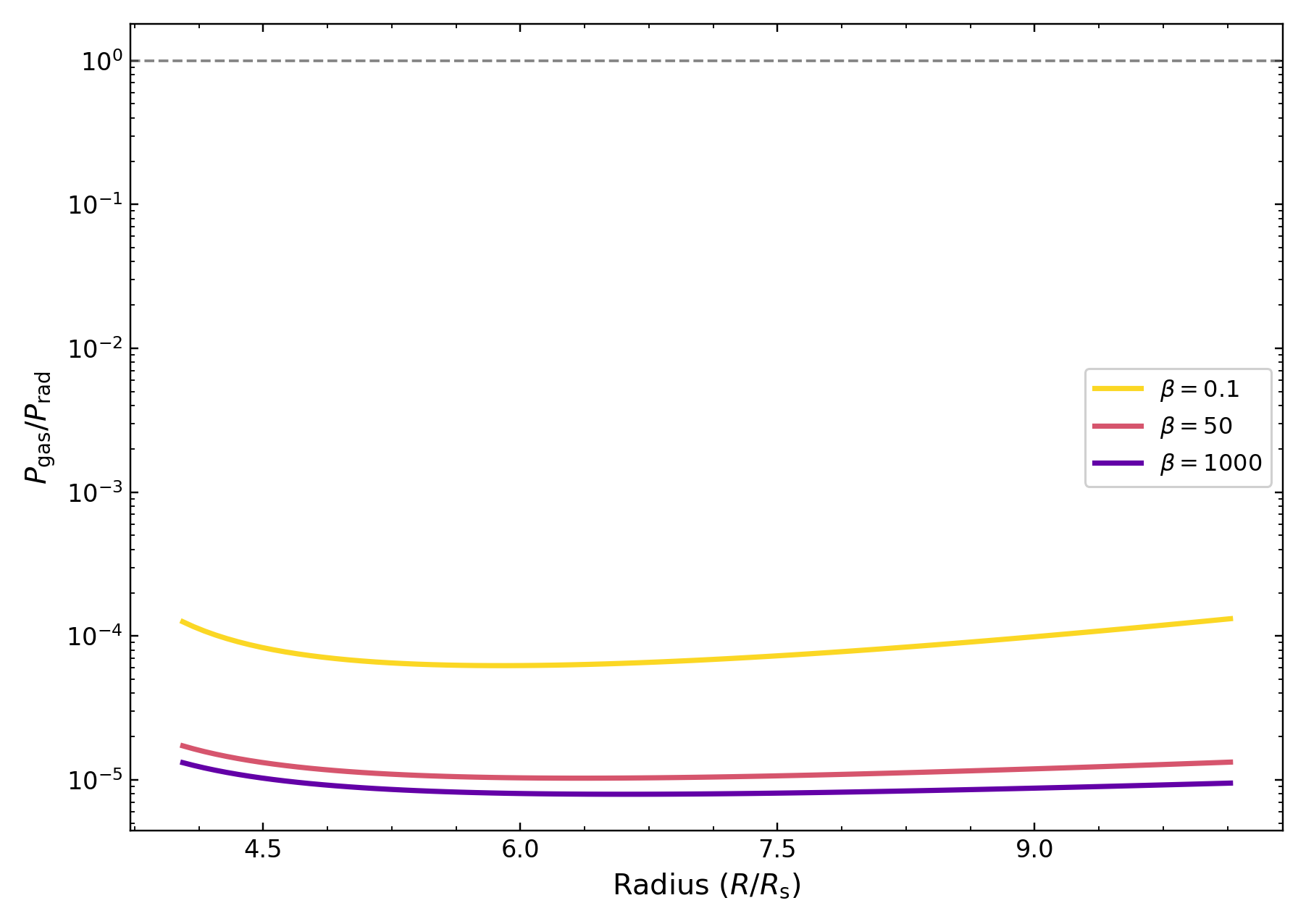}
\caption{Pressure-ratio profiles for $M_{\rm BH}=10^7\,\msun$. Left: representative accretion rates at fixed $\beta=50$. Right: representative magnetic parameters at fixed $\mdotd=1$. The dashed line marks $P_{\rm gas}=P_{\rm rad}$.}
\label{fig:app_mdot_pgas_prad_1e7}
\label{fig:app_beta_pgas_prad_1e7}
\label{lastpage}
\end{figure*}
\clearpage

\end{document}